&pdflatex
\documentclass[3p]{elsarticle}
\usepackage{url}
\usepackage{amssymb}
\usepackage{amsmath}
\usepackage{amsthm}
\usepackage{nicefrac}
\usepackage{tabu}
\usepackage{algpseudocode}
\usepackage[section]{algorithm}

\begin{document}

%% fix left\right spacing issues?
\let\originalleft\left
\let\originalright\right
\renewcommand{\left}{\mathopen{}\mathclose\bgroup\originalleft}
\renewcommand{\right}{\aftergroup\egroup\originalright}

%% symbols
\newcommand{\cpp}{{\nolinebreak C\texttt{++} }}

\newcommand{\BigO}[1]{\mathop{}\!O{\left(#1\right)}}

\newcommand{\Del}[1]{\operatorname{Del}\left(#1\right)}
\newcommand{\Vor}[1]{\operatorname{Vor}\left(#1\right)}
\newcommand{\Conv}[1]{\operatorname{Conv}\left(#1\right)}
\newcommand{\DelC}[1]{\operatorname{Del}|_{\Gamma}\left(#1\right)}
\newcommand{\DelS}[1]{\operatorname{Del}|_{\Sigma}\left(#1\right)}
\newcommand{\DelV}[1]{\operatorname{Del}|_{\Omega}\left(#1\right)}
\newcommand{\TC}{\mathcal{T}|_{\Gamma}}
\newcommand{\TS}{\mathcal{T}|_{\Sigma}}
\newcommand{\TV}{\mathcal{T}|_{\Omega}}

\newcommand{\reledge}{h_{r}}

\newtheorem{remark}{Remark}[section]

\renewcommand{\topfraction}{.85}
\renewcommand{\bottomfraction}{.75}
\renewcommand{\textfraction}{.15}
\renewcommand{\floatpagefraction}{.66}
\renewcommand{\dbltopfraction}{.66}
\renewcommand{\dblfloatpagefraction}{.66}

\begin{frontmatter}

%\dochead{25th International Meshing Roundtable (IMR25)}

\title{Conforming restricted Delaunay mesh generation for piecewise smooth complexes\tnoteref{tnote1}}

\tnotetext[tnote1]{To appear at the 25th International Meshing Roundtable.}

\author[a]{Darren Engwirda\corref{cor1}}
\ead{engwirda@mit.edu}

\address[a]{Department of Earth, Atmospheric and Planetary Sciences, Massachusetts Institute of Technology, 54-918, 77 Massachusetts Avenue, Cambridge, MA 02139-4307, USA}

\begin{abstract}
A Frontal-Delaunay refinement algorithm for mesh generation in piecewise smooth domains is described. Built using a restricted Delaunay framework, this new algorithm combines a number of novel features, including: (i) an unweighted, conforming restricted Delaunay representation for domains specified as a (non-manifold) collection of piecewise smooth surface patches and curve segments, (ii) a protection strategy for domains containing curve segments that subtend sharply acute angles, and (iii) a new class of off-centre refinement rules designed to achieve high-quality point-placement along embedded curve features. Experimental comparisons show that the new Frontal-Delaunay algorithm outperforms a classical (statically weighted) restricted Delaunay-refinement technique for a number of three-dimensional benchmark problems.
\end{abstract}

\cortext[cor1]{Corresponding author. Tel.: +1-212-678-5521}
\end{frontmatter}

%\email{engwirda@mit.edu}

\section{Introduction}
\label{section_introduction}

Mesh generation is a key component in a variety of mathematical modelling and simulation tasks, including problems in computational engineering, numerical modelling, and computer graphics and animation. Given a general volumetric domain, described by a network of curves $\Gamma\subset\mathbb{R}^{3}$, a collection of surfaces $\Sigma\subset\mathbb{R}^{3}$ and an enclosed volume $\Omega\subset\mathbb{R}^{3}$, the three-dimensional meshing problem consists of tessellating $\Gamma$, $\Sigma$ and $\Omega$ into a \textit{mesh} of non-overlapping \textit{simplexes} (edges, triangles and tetrahedrons), such that all geometrical, topological and user-defined constraints are satisfied. While some input domains can be described in terms of smooth entities, it is typical to deal with objects that are only \textit{piecewise-smooth}, consisting of locally manifold surface patches that meet at sharp 0- or 1-dimensional features. Additionally, input domains can incorporate so-called \textit{free} curve and vertex constraints, comprising sets of entities unconnected to the bounding surface patches $\Sigma$. In this study, a new Delaunay-refinement type algorithm is presented to construct meshes for piecewise smooth domains -- forming a Delaunay tetrahedralisation that includes a subset of \textit{restricted} edges, triangles and tetrahedrons that provide provably-good topological and geometrical approximations to the input curve network $\Gamma$, surface structure $\Sigma$ and enclosed volume $\Omega$. Using a new class of \textit{off-centre} refinement rules, the proposed algorithm is cast as a \textit{Frontal-Delaunay} scheme -- seeking to extend the surface- and volume-meshing techniques presented by the author in \cite{Engwirda2016157,Engwirda2015330,engwirda2014locally}. Compared to existing restricted Delaunay-refinement techniques, the methods described here incorporate a number of novel features, including: (i) the use of a conforming, \textit{unweighted} restricted Delaunay representation for piecewise smooth geometries, (ii) the development of a new \textit{collar-based} method to protect sharply acute features present in the input geometry, and (iii) the use of off-centre point-placement rules to refine curve, surface and volumetric elements in a provably-good manner.

%\smallskip

%\begin{nomenclature}
%\begin{deflist}[AAAA]

\subsection{Nomenclature}

\medskip

The following work is based on the \textit{restricted} Delaunay framework. The reader is referred to \cite{ChengDeyShewchuk} for formal definitions, discussions and proofs.

\medskip

\begin{figure}[t]
\centering

{
\footnotesize
\tabulinesep=0pt

\smallskip

\begin{tabu}{ccc}

\begin{minipage}[c]{0.275\textwidth}
\centering
\includegraphics[width=3.75cm]{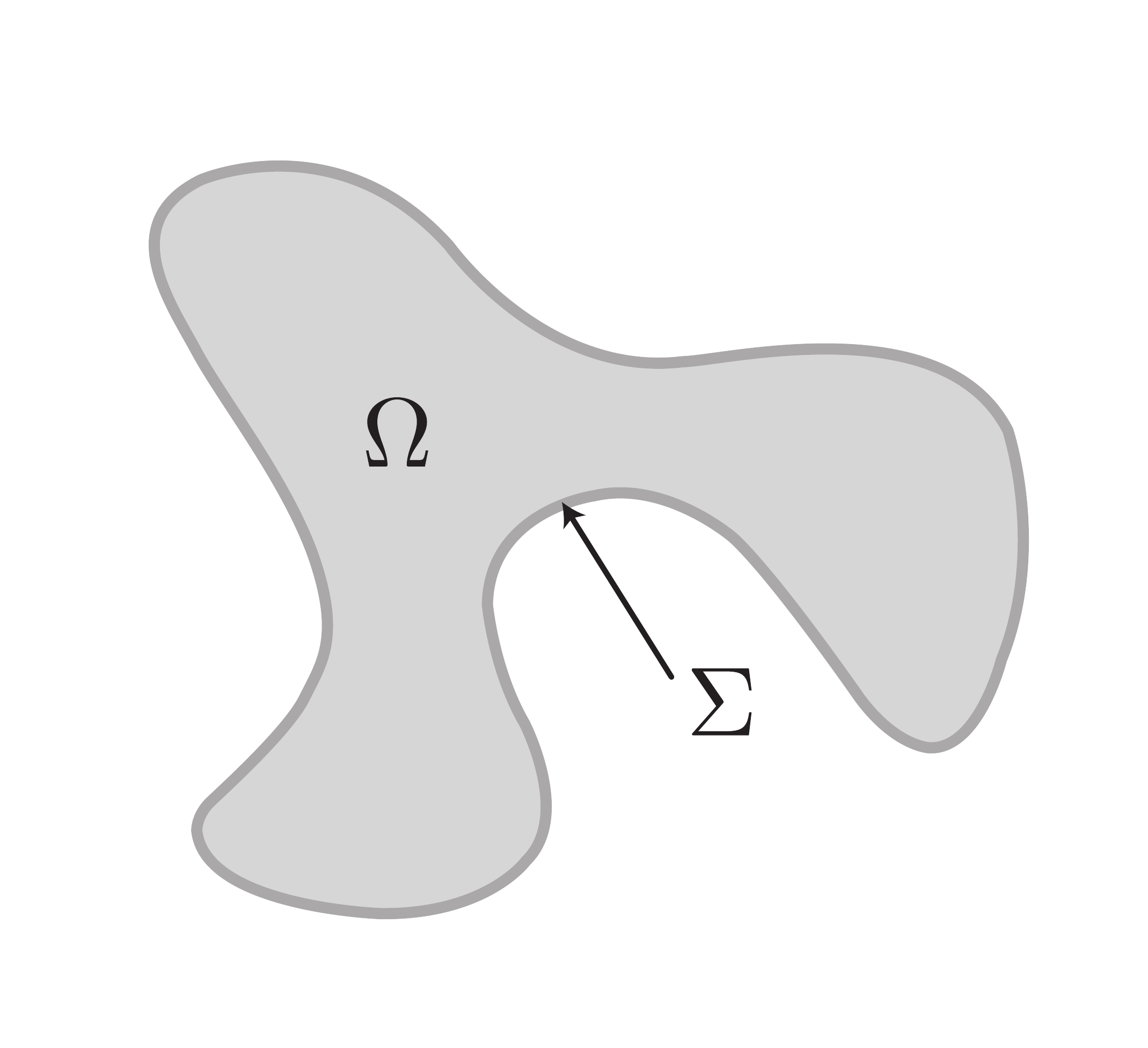} 
\end{minipage} &
\begin{minipage}[c]{0.275\textwidth}
\centering
\includegraphics[width=3.75cm]{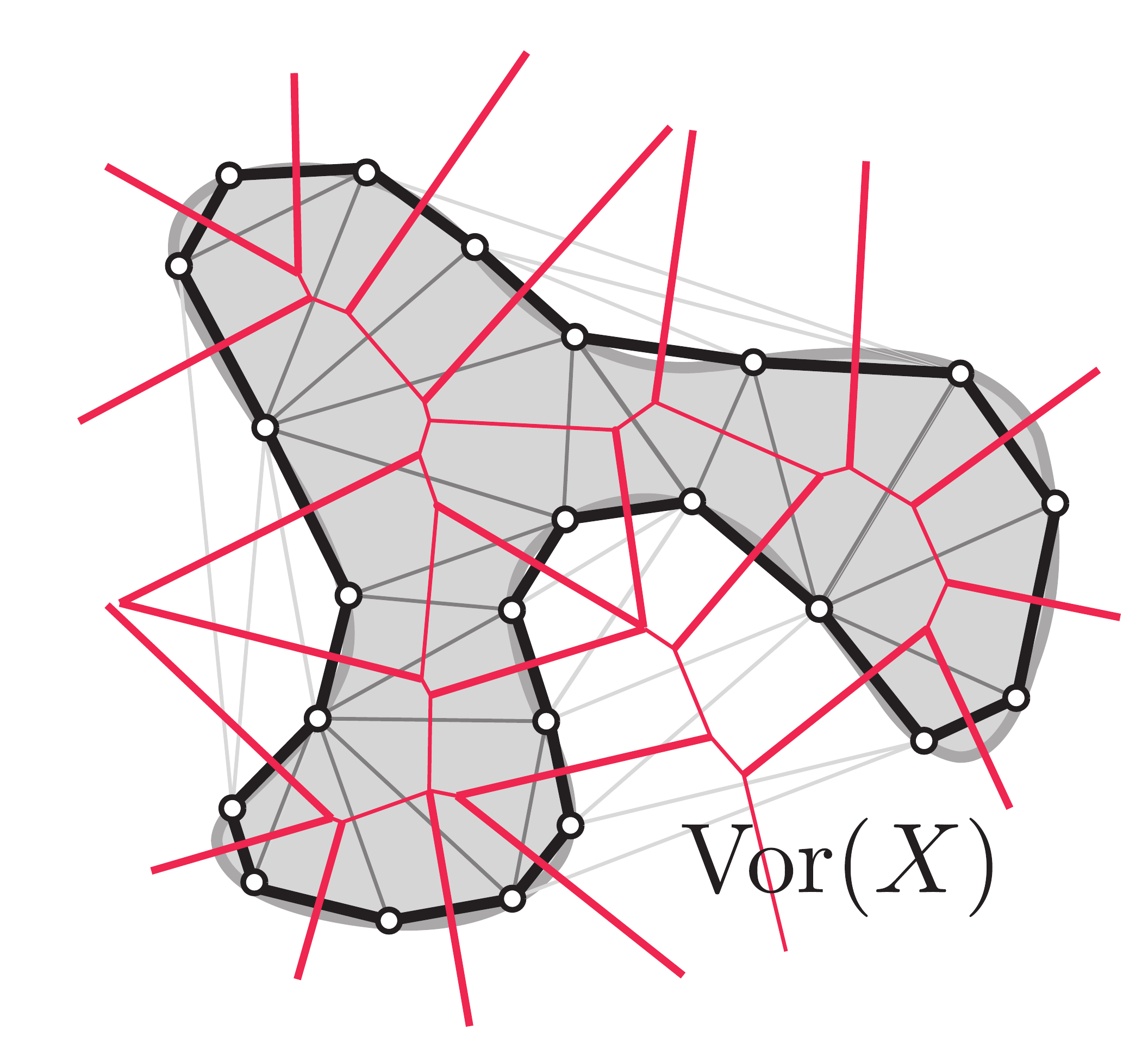} 
\end{minipage} &
\begin{minipage}[c]{0.275\textwidth}
\centering
\includegraphics[width=3.75cm]{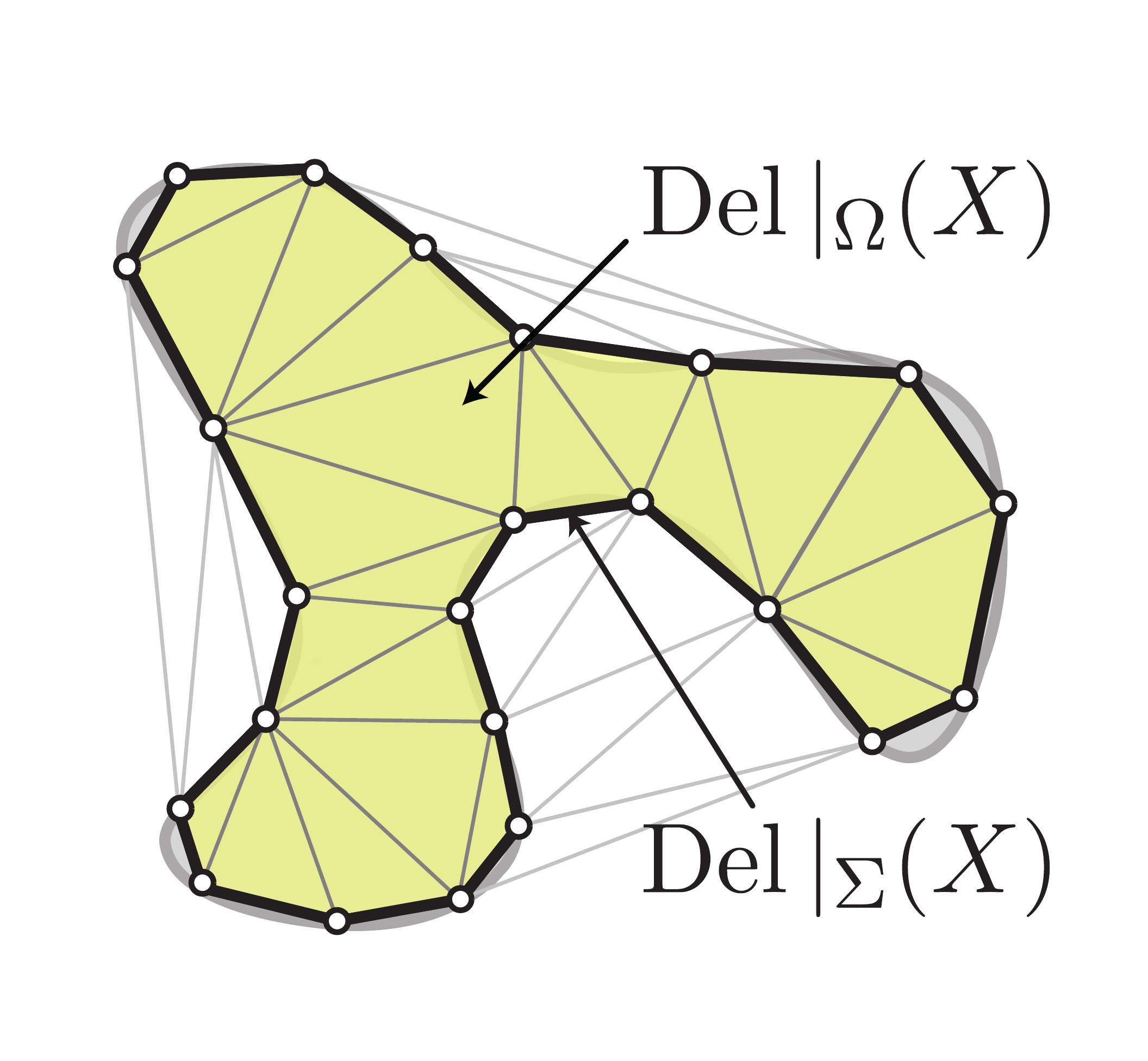} 
\end{minipage} \\

(i) & (ii) & (iii) 

\end{tabu}}

\caption{Restricted Delaunay tessellations for a smooth domain in $\mathbb{R}^{2}$, showing (i) the bounding contour $\Sigma$ and enclosed interior $\Omega$, (ii) the Delaunay tessellation $\Del{X}$ and Voronoi diagram $\Vor{X}$, and (iii) the restricted boundary and interior tessellations $\DelS{X}$ and $\DelV{X}$. In three-dimensions, the restricted surface triangulation $\DelS{X}$ is a triangular complex that covers the surface $\Sigma$. The restricted volume tessellation $\DelV{X}$ is a tetrahedral complex that fills the volume $\Omega$. }

\label{figure_restricted_delaunay}

\end{figure}

{\raggedleft
\begin{minipage}{0.950\textwidth}
\begin{itemize} \itemsep 0pt

\item{$X$}: {A set of points in $\mathbb{R}^{3}$, associated with the tessellation.}

\item{$\Del{X}$}: {The Delaunay triangulation of the points $X$.}

\item{$\Vor{X}$}: {The Voronoi complex associated with the points $X$.}

\item{$\Gamma, \Sigma, \Omega$}: {The input geometry: a collection of curve segments, surface patches and volumes embedded in $\mathbb{R}^{3}$.}

\item{$\DelC{X}$}: {A Delaunay sub-complex $\DelC{X}\subseteq\Del{X}$, restricted to the curve network $\Gamma$. $\DelC{X}$ contains any 1-simplex $e\in\Del{X}$ whose dual Voronoi face $v_{f}\subseteq\Vor{X}$ intersects $\Gamma$ (See Figure~\ref{figure_restricted_delaunay}(ii)--\ref{figure_restricted_delaunay}(iii)).}

\item{$\DelS{X}$}: {A Delaunay sub-complex $\DelS{X}\subseteq\Del{X}$, restricted to the surfaces patches $\Sigma$. $\DelS{X}$ contains any 2-simplex $f\in\Del{X}$ whose dual Voronoi edge $v_{e}\subseteq\Vor{X}$ intersects $\Sigma$ (See Figure~\ref{figure_restricted_delaunay}(ii)--\ref{figure_restricted_delaunay}(iii)).}

\item{$\DelV{X}$}: {A Delaunay sub-complex $\DelV{X}\subseteq\Del{X}$, restricted to the interior volumes $\Omega$. $\DelV{X}$ contains any 3-simplex $\tau\in\Del{X}$ with a dual Voronoi vertex $v_{x}\subseteq\Vor{X}$ interior to $\Omega$ (See Figure~\ref{figure_restricted_delaunay}(ii)--\ref{figure_restricted_delaunay}(iii)).}

\item{$\rho_{d}(\tau)$}: {The radius-edge ratio associated with a $d$-simplex $\tau$. Defined as the ratio of the radius of the circumball of $\tau$ to the length of its shortest edge.}

\item{$\epsilon_{1}(e)$}: {The surface discretisation error associated with a 1-simplex $e\in\DelC{X}$. Defined as the length from the centre of $\operatorname{SDB}_{1}(e)$ to the centre of the diametric ball of $e$.}

\item{$\epsilon_{2}(f)$}: {The surface discretisation error associated with a 2-simplex $f\in\DelS{X}$. Defined as the length from the centre of $\operatorname{SDB}_{2}(f)$ to the centre of the diametric ball of $f$.}

\item{$\operatorname{SDB}_{1}(e)$}: {The surface Delaunay ball $B(\mathbf{c}_{e},r)$ associated with a 1-simplex $e\in\DelC{X}$. Balls are centred at intersections between the Voronoi faces $v_{f}\in\Vor{X}$ and the curve network $\Gamma$, such that $\mathbf{c}_{e}=v_{f}\cap\Gamma$.}

\item{$\operatorname{SDB}_{2}(f)$}: {The surface Delaunay ball $B(\mathbf{c}_{f},r)$ associated with a 2-simplex $f\in\DelS{X}$. Balls are centred at intersections between the Voronoi edges $v_{e}\in\Vor{X}$ and the surface patches $\Sigma$, such that $\mathbf{c}_{f}=v_{e}\cap\Sigma$.}

\item{$\bar{h}(\mathbf{x})$}: {The mesh-size function. A function $f(\mathbf{x}): \mathbb{R}^{3}\rightarrow\mathbb{R}^{+}$ defining the target edge length at points $\mathbf{x}\in\Omega$.}

%\defitem{$a(f)$}\defterm{The area-length ratio associated with a given triangle $f$. Defined as $a(f)=\nicefrac{A}{\|\mathbf{e}\|^{2}_{\text{rms}}}$, where $A$ is the signed area of $f$ and $\|\mathbf{e}\|_{\text{rms}}$ is the root-mean-square edge length. The area-length ratio is a consistent scalar measure of triangular element quality.}

\item{$v(\tau)$, $a(f)$}: {The volume-length and area-length ratios associated with a given tetrahedron $\tau$ or triangle $f$. Defined as $v(\tau)=\nicefrac{V}{\|\mathbf{e}\|^{3}_{\text{rms}}}$ and $a(f)=\nicefrac{A}{\|\mathbf{e}\|^{2}_{\text{rms}}}$, where $V$ is the signed volume of $\tau$, $A$ is the signed area of $f$ and $\|\mathbf{e}\|_{\text{rms}}$ is the root-mean-square edge length. The volume-length and area-length ratios are \textit{robust} measures of tetrahedral and triangular element quality.}

\end{itemize}
\end{minipage}
}

%\end{deflist}
%\end{nomenclature}
%\vspace*{6pt}

%The reader is referred to \cite{ChengDeyShewchuk} for formal definitions, discussions and proofs.

\subsection{Preliminaries}

\medskip

%Delaunay-refinement schemes are \textit{top-down} algorithms -- based on the incremental refinement of a bounding Delaunay tessellation.

Many successful three-dimensional meshing algorithms employ \textit{Delaunay-based} strategies \cite{Chew89Provable,Ruppert93Provable,Ruppert95Provable,Shewchuk97PhD,Shewchuk98Tetra,
Cheng03WeightedDelaunay,Cheng10PiecewiseSmoothMeshing,boissonnat03ProvablyGoodSurface,
boissonnat05ProvablyGoodMeshing,jamin2013cgalmesh}, based on the progressive refinement of a coarse initial Delaunay triangulation that spans the input geometry. At each step, elements that violate a set of constraints are identified and the worst offending elements are \textit{eliminated}. Elimination is achieved through the insertion of additional \textit{Steiner-vertices} located at the so-called \textit{refinement-points} associated with the elements in question. Delaunay-refinement algorithms have been developed for planar \cite{Chew89Provable,Ruppert93Provable,Ruppert95Provable}, surface \cite{boissonnat03ProvablyGoodSurface,boissonnat05ProvablyGoodMeshing} and volumetric domains \cite{Shewchuk98Tetra,Si10ConstrainedTetra,Si:2015:TDQ:2732672.2629697}. The reader is referred to \cite{ChengDeyShewchuk} for additional information and summary.

This study is focused on use of the so-called \textit{restricted} Delaunay methodology \cite{Edelsbrunner97Restricted,Cheng10PiecewiseSmoothMeshing,jamin2013cgalmesh} to provide a framework for the approximation of 1-, 2- and 3-dimensional topological features via Delaunay sub-complexes. Such techniques have been the focus of previous work, including, for example \cite{boissonnat03ProvablyGoodSurface,boissonnat05ProvablyGoodMeshing,
Cheng10PiecewiseSmoothMeshing,ChengDeyShewchuk} and previous studies by the author in \cite{Engwirda2016157,Engwirda2015330,engwirda2014locally}, where it has been shown that various geometrical and topological guarantees of fidelity are achieved through a careful sampling of the geometrical inputs. Compared to other approaches, the restricted Delaunay framework incorporates a number of desirable characteristics, chiefly: (i) the ability to sample curve-, surface- and volumetric-features in a unified manner, and (ii) the development of \textit{geometry-agnostic} meshing algorithms. These characteristics are useful from both a theoretical and software development standpoint: the use of a unified meshing framework obviates non-trivial difficulties associated with the construction of \textit{constrained} Delaunay complexes that conform to curve and surface constraints \cite{Shewchuk08CDTTheory,Si08ConstrainedTetra}, while the use of a geometry-agnostic formulation facilitates the development of meshing software that supports a broad class of input geometry types and definitions.

Consistent with previous work by the author \cite{Engwirda2016157,Engwirda2015330,engwirda2014locally}, the present study combines the restricted Delaunay framework with a so-called \textit{Frontal-Delaunay} methodology -- seeking to achieve very high-quality Delaunay-based mesh generation through use of a hybrid, advancing-front type strategy. Using an appropriate set of \textit{off-centre} point-placement rules, the Frontal-Delaunay approach aims to combine the best features of classical Delaunay-refinement and advancing-front type techniques, leading to high-quality Delaunay meshes that satisfy a theoretical bounds and guarantees. It is expected that this algorithm may be of interest to users who place a high premium on mesh quality, including those operating in the areas of computational engineering and numerical simulation.

The present study is organised as follows: an overview of the restricted Delaunay framework is presented in Section~\ref{section_restricted_tria}, including a detailed discussion of the techniques used to recover restricted Delaunay edges, triangles and tetrahedrons. A hierarchical restricted Delaunay-refinement algorithm is presented in Section~\ref{section_restricted_refinement}, with a new class of curve-based off-centre refinement rules described in Section~\ref{section_offcentres}. A technique for the protection of acute features is presented in Section~\ref{section_protecting_angles}, allowing domains containing curve features that subtend arbitrarily small angles to be meshed. Comparisons between a conventional (statically-weighted) restricted Delaunay-refinement algorithm and the proposed Frontal-Delaunay scheme is presented in Section~\ref{section_results}, contrasting output quality and computational performance.

\section{Restricted Delaunay Edges, Triangles \& Tetrahedrons}
\label{section_restricted_tria}

The meshing algorithms presented in this study are based on the \textit{restricted} Delaunay paradigm -- a framework utilising a hierarchy of Delaunay sub-complexes to provide consistent and conforming approximations to embedded geometrical features. In the context of three-dimensional meshes, the bounding Delaunay tessellation $\Del{X}$ is a tetrahedral complex -- of sufficient size to enclose the input domain. Embedded within $\Del{X}$ are a set of \textit{restricted} Delaunay sub-complexes: $\DelC{X}$, $\DelS{X}$ and $\DelV{X}$, providing discrete approximations to the curve network $\Gamma$, surface structure $\Sigma$ and interior volumes $\Omega$, respectively. The restricted curve complex $\DelC{X}\subseteq\Del{X}$ contains the set of 1-simplexes $e\in\Del{X}$ that provide \textit{good} piecewise linear approximations to the curve network $\Gamma$. Similarly, the restricted surface and volume complexes $\DelS{X}\subseteq\Del{X}$ and $\DelV{X}\subseteq\Del{X}$ contain the sets of 2- and 3-simplexes $f\in\Del{X}$ and $\tau\in\Del{X}$ that provide good approximations to the surface patches $\Sigma$ and volumes $\Omega$. An overview of these concepts is provided in, for example \cite{ChengDeyShewchuk,boissonnat03ProvablyGoodSurface,
boissonnat05ProvablyGoodMeshing,jamin2013cgalmesh,Cheng10PiecewiseSmoothMeshing}.

\begin{figure}[t]
\centering

{
\footnotesize
\tabulinesep=0pt

\smallskip

\begin{tabu}{ccc}

\begin{minipage}[c]{0.275\textwidth}
\centering
\includegraphics[width=4.25cm]{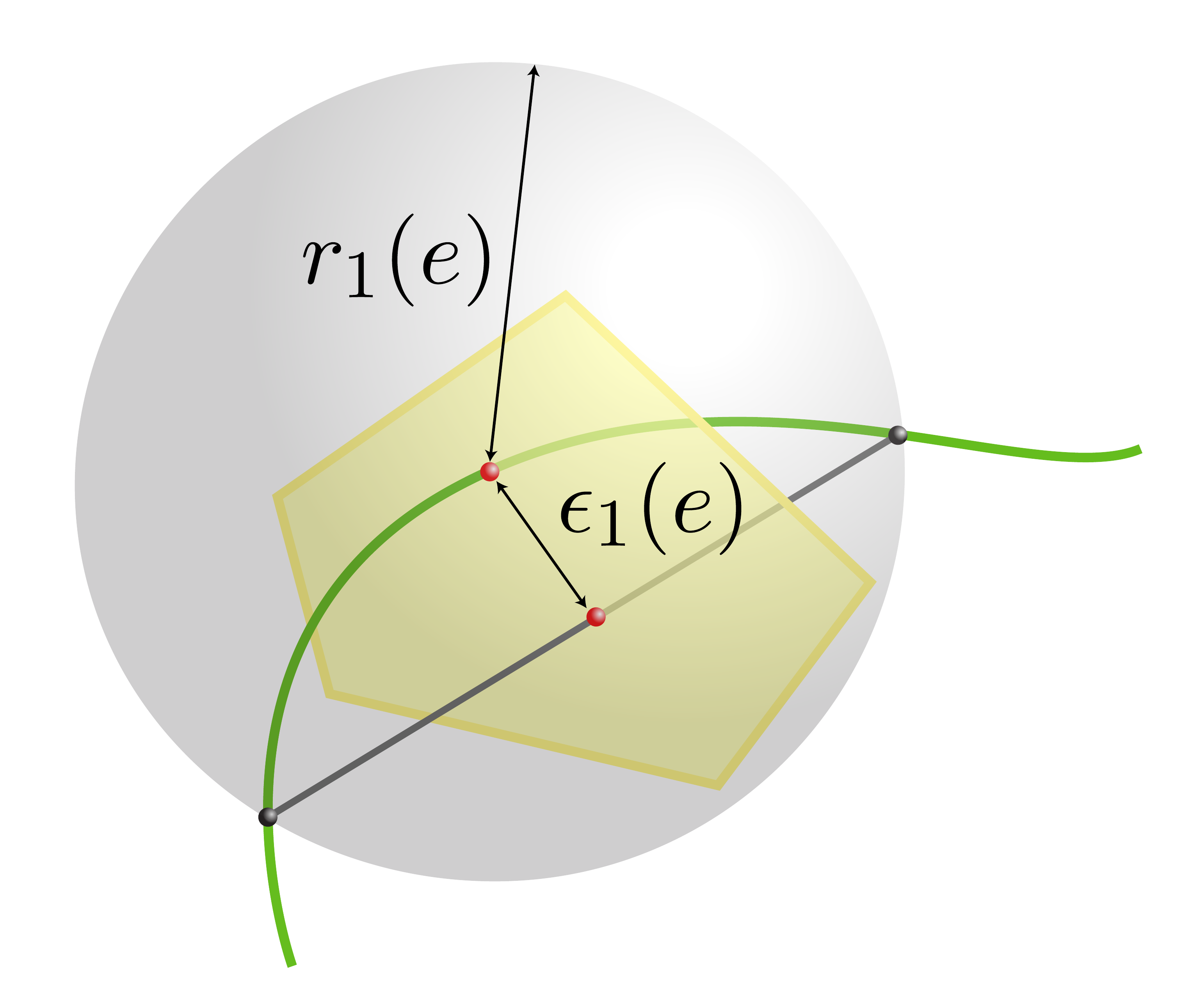}
\end{minipage} &
\begin{minipage}[c]{0.275\textwidth}
\centering
\includegraphics[width=4.25cm]{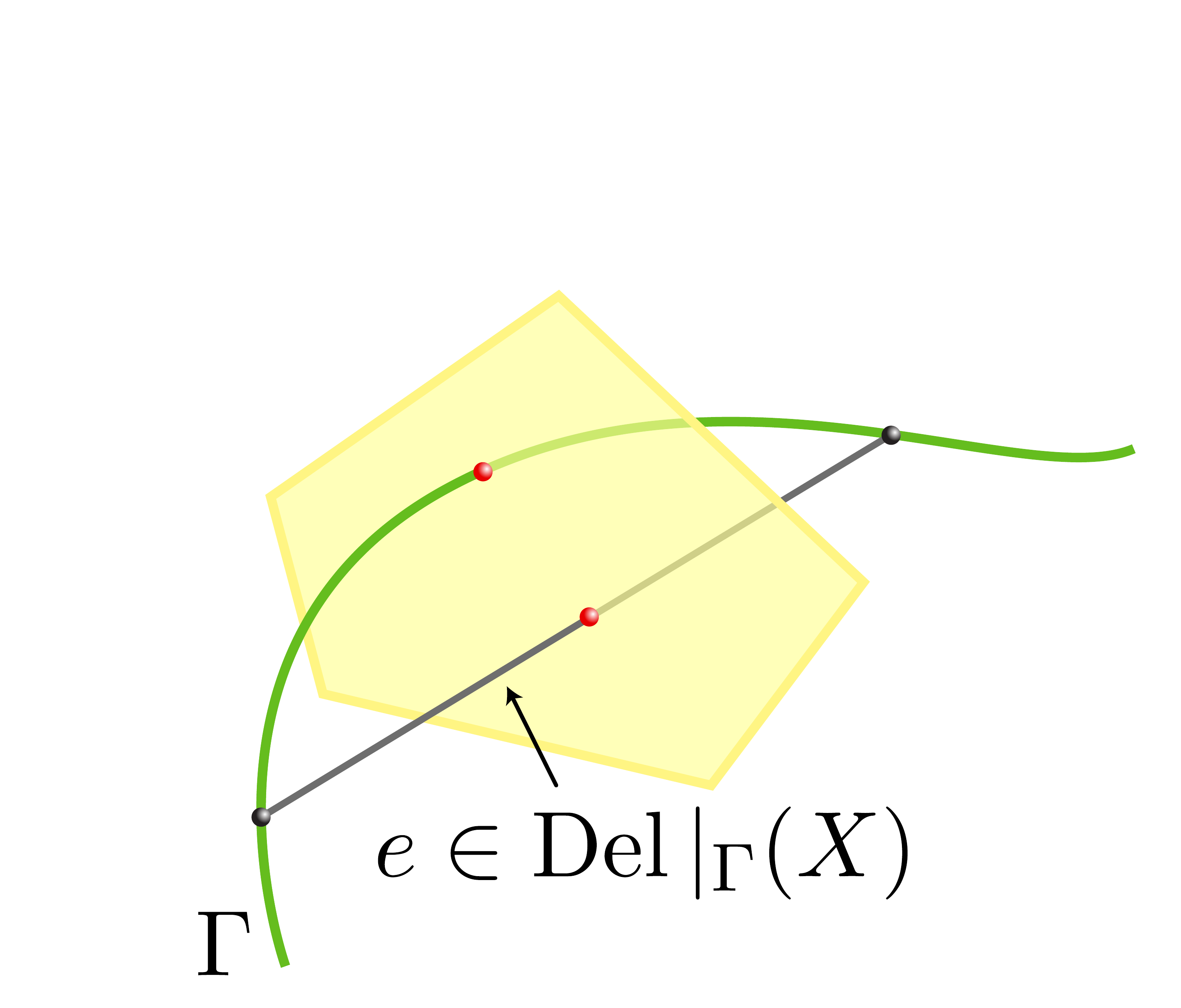}
\end{minipage} &
\begin{minipage}[c]{0.275\textwidth}
\centering
\includegraphics[width=4.25cm]{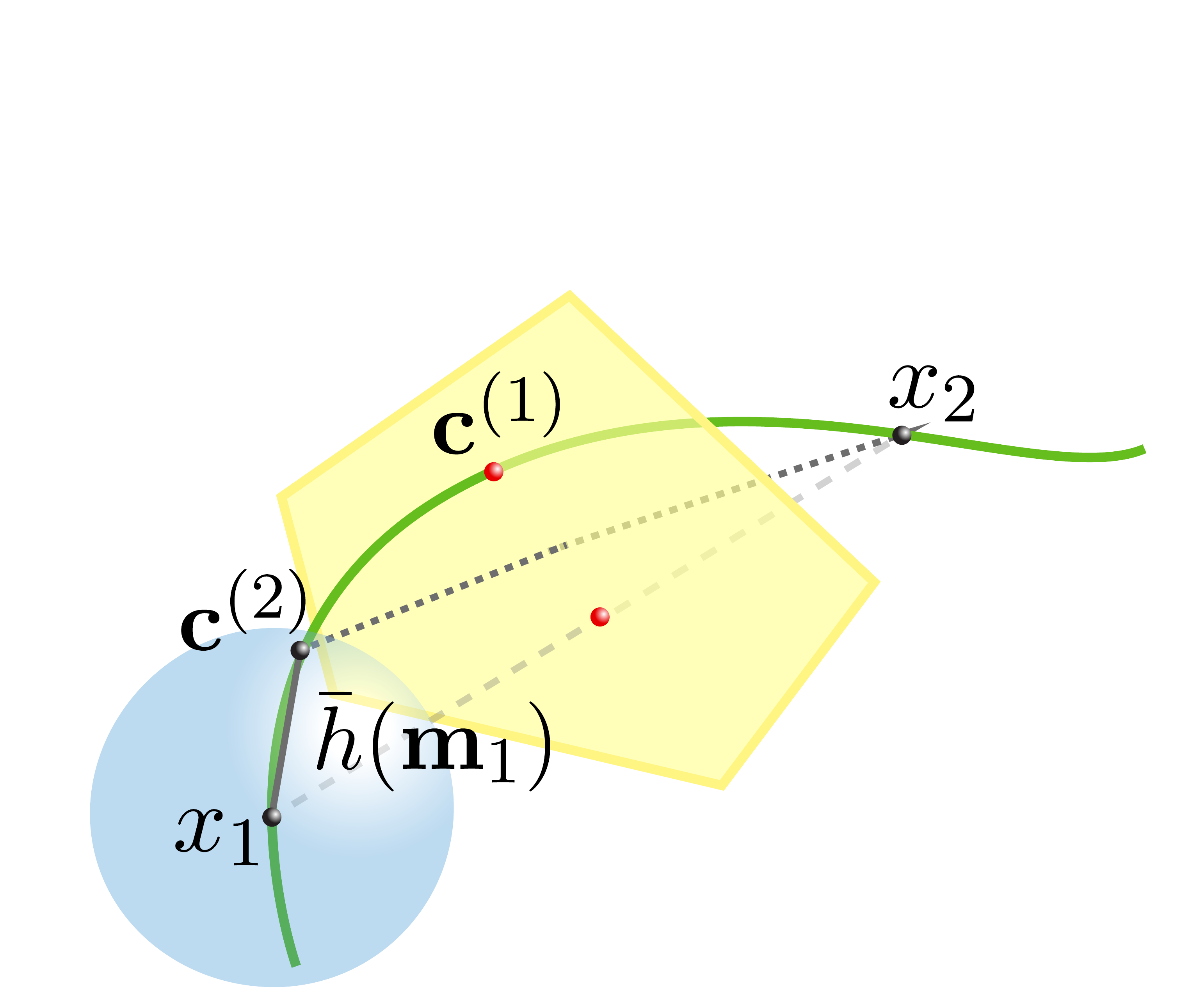}
\end{minipage} 

\\

\smallskip

(i) & (ii) & (iii) 

\end{tabu}}

\caption{A restricted Delaunay 1-simplex $e\in\DelC{X}$ associated with a curve segment $\Gamma$, showing (i) the intersection of the associated Voronoi face $v_{f}\subseteq\Vor{X}$ and the curve network $\Gamma$, (ii) the associated Surface Delaunay Ball $\operatorname{SDB}(e)$, where $r_{1}(e)$ denotes the SDB radius and $\epsilon_{1}(e)$ the surface discretisation error, and (iii) the off-centre type refinement rule, showing the placement of a locally size-optimal point $\mathbf{c}^{(2)}$ about a frontal vertex $\mathbf{x}_{1}$. }

\label{figure_restricted_edge}

\end{figure}

Restricted Delaunay techniques exploit the duality between Delaunay tessellations and Voronoi complexes -- using such considerations to compute membership for the restricted Delaunay sub-complexes $\DelC{X}$, $\DelS{X}$ and $\DelV{X}$. Specifically, $\DelV{X}$ contains any 3-simplex $\tau_{i}$ that is associated with an internal Voronoi vertex $\mathbf{v}_{i}\in\Omega$, while $\DelS{X}$ contains any 2-simplex $f_{j}$ associated with a Voronoi segment $\mathbf{v}_{ab}\in\Vor{X}$ that intersects the surface structure $\Sigma$, such that $\mathbf{v}_{ab}\cap\Sigma \neq \emptyset$. These are well-known results. Less widely utilised is a mechanism for identifying the restricted 1-simplexes embedded in a tetrahedral complex. In \cite{Rineau2008}, Rineau and Yvinec present a methodology based on dual \textit{Voronoi-faces}. Specifically, $\DelC{X}$ contains any 1-simplex $e_{k}$ associated with a Voronoi face $v_{f}\subseteq\Vor{X}$ that intersects the curve network $\Gamma$, such that $v_{f}\cap\Gamma \neq \emptyset$. Note that, by definition, the faces of the Voronoi complex $v_{f}$ are convex polygons, oriented normally to their associated Delaunay edges. See Figure~\ref{figure_restricted_edge}(i) for details. 

Each element in a restricted Delaunay sub-complex is also associated with a circumscribing ball. For tetrahedrons in $\DelV{X}$ such balls are unique -- being equivalent to the set of circumscribing spheres that pass through the vertices associated with each element. For edges in $\DelC{X}$ and triangles in $\DelS{X}$, each element is instead associated with a so-called \textit{Surface Delaunay Ball} $\operatorname{SDB}(f_{i})$. These balls are the circumscribing spheres centred upon intersections of the associated Voronoi dual with the input geometry. In the case of multiple intersections, the corresponding ball of maximum radius is selected. See Figure~\ref{figure_restricted_edge}(ii) for details. Surface Delaunay Balls also support a discrete measure of geometrical fidelity. Specifically, given an edge or surface element $e_{k}\in\DelC{X}$ or $f_{j}\in\DelS{X}$, the distance between the centre of the associated diametric ball and surface ball is a one-sided Hausdorff metric -- a measure of the geometrical approximation error induced by the piecewise linear Delaunay mesh.

\section{A Restricted Delaunay-refinement Algorithm}
\label{section_restricted_refinement}

An algorithm for the meshing of piecewise smooth complexes embedded in $\mathbb{R}^3$ is presented here, as an extension of previous work by the author in \cite{Engwirda2016157,Engwirda2015330,engwirda2014locally} and by various other authors, including: Rineau and Yvinec \cite{Rineau2008}, Cheng, Dey and Shewchuk \cite{ChengDeyShewchuk}, Cheng, Dey and Levine \cite{cheng2008practical}, and Oudot, Rineaua and Yvinec \cite{oudot2005meshing}. This method is related to the \texttt{CGALMESH} algorithm -- a \textit{classical} restricted Delaunay-refinement approach available as part of the \texttt{CGAL} library \cite{cgal:pt-t3-15a}, and summarised by Jamin, Alliez, Yvinec and Boissonnat in \cite{jamin2013cgalmesh}. The algorithm presented here differs significantly in the methodology used for the recovery of 1-dimensional features. Specifically, in the current work, curve constraints are represented as an (unweighted) conforming restricted Delaunay sub-complex, as outlined in Section~\ref{section_restricted_tria}, and consistent with the techniques described by Rineau and Yvinec in \cite{Rineau2008}. The \texttt{CGALMESH} algorithm is instead based on a so-called \textit{protecting-balls} strategy \cite{ChengDeyShewchuk,cheng2008practical}, in which a static discretisation of the curve network $\Gamma$ is conducted as an initialisation step, with the resulting edge segments protected by a \textit{suitably-weighted} Delaunay tessellation. Such a strategy preserves edge constraints throughout the subsequent surface- and volume-refinement iterations, though it does not guarantee that element quality thresholds are satisfied in the neighbourhood of such constraints. In some cases, this behaviour can lead to the creation of lower-quality elements adjacent to 1-dimensional features. %Further comparisons are made in Section~\ref{section_results}. 

\subsection{Preliminaries}

\medskip

As per Jamin et~al.~\cite{jamin2013cgalmesh}, the development of restricted Del\-aunay-refinement algorithms is \textit{geometry-agnostic}, being independent of the specific definition of the underlying geometry inputs. It is required only that the framework support a set of so-called \textit{oracle} predicates, used to compute: (i) the intersection of convex polygons (Voronoi faces) with the curve network $\Gamma$ (ii) the intersection of line segments (Voronoi edges) with the surface patches $\Sigma$, and (iii) the intersection of points (Voronoi vertices) with the enclosed volume $\Omega$. The Frontal-Delaunay algorithm presented in subsequent sections additionally requires the computation of intersections between the curves $\Gamma$ and surfaces $\Sigma$ with spheres and oriented disks. While a broad class of geometry descriptions are supported at the theoretical level, in this study, attention is restricted to the development of so-called \textit{re-meshing} operations, in which the input geometries are specified in terms of discrete polylines and triangulated surfaces $\mathcal{P}$. This restriction is made to facilitate the construction of simple oracle predicates. Future work is intended to focus on the development of predicates for more general descriptions, including domains defined by implicit, parametric and analytic functions.

Following Jamin et~al.~\cite{jamin2013cgalmesh}, the Delaunay-refinement algorithm takes as input a volumetric domain $\Omega$, described by an enclosing (possibly non-manifold) surface $\Sigma\subseteq\mathbb{R}^{3}$, a network of curve segments $\Gamma\subseteq\mathbb{R}^{3}$, an upper bound on the allowable element radius-edge ratio $\bar{\rho}$, a mesh size function $\bar{h}\left(\mathbf{x}\right)$ defined at all points spanned by the domain, and an upper bound on the allowable surface discretisation error $\bar{\epsilon}(\mathbf{x})$. The algorithm returns a discretisation $\TC$ of the curve network $\Gamma$, a triangulation $\TS$ of the surface patches $\Sigma$, and a triangulation $\TV$ of the enclosed volume $\Omega$. Here $\TC$, $\TS$ and $\TV$ are restricted Delaunay sub-complexes, such that $\TC=\DelC{X}$, $\TS=\DelS{X}$ and $\TV=\DelV{X}$. Note that $\DelC{X}$ is an edge complex, $\DelS{X}$ is a triangular complex, and $\DelV{X}$ and $\Del{X}$ are tetrahedral complexes. The Delaunay-refinement algorithm is summarised in Algorithm~\ref{algorithm_restricted_delaunay_refinement}.

The Delaunay-refinement algorithm is designed to provide a number of geometrical and topological guarantees on the output mesh, specifically: (i) that all elements in the volumetric tessellation $\tau\in\TV$ satisfy constraints on both the element shape and size, such that $\rho(\tau)\leq\bar{\rho}$, and $h(\tau)\leq \bar{h}\left(\mathbf{x}_{\tau}\right)$, (ii) that all elements in the embedded surface triangulation $f\in\TS$ are guaranteed to satisfy similar element shape and size constraints, in addition to an upper bound on the allowable surface discretisation error, such that $\epsilon(f)\leq\bar{\epsilon}(\mathbf{x}_{f})$, (iii) that the surface triangulation $\TS$ is \textit{topologically-consistent}, ensuring that $\DelS{X}$ is uniformly 2-manifold in its interior, and is consistent with the input geometry at non-manifold features, (iv) that all elements in the embedded curve triangulation $e\in\TC$ are guaranteed to satisfy similar element size and surface error constraints, and (v) that the curve discretisation $\TC$ is also topologically-consistent, ensuring that $\DelC{X}$ is uniformly 1-manifold in its interior, and is consistent with the input geometry at non-manifold features. Making use of properties of the restricted Delaunay tessellation \cite{Edelsbrunner97Restricted}, it is known that the triangulations $\TC$, $\TS$ and $\TV$ are good piecewise linear approximations to the input curves $\Gamma$, surfaces $\Sigma$ and volumes $\Omega$, provided that the magnitude of the mesh-size function $\bar{h}\left(\mathbf{x}\right)$ is sufficiently small. Under such conditions it is known that the triangulations $\TC$, $\TS$ and $\TV$ are homeomorphic to the underlying curve, surface and volume definitions $\Gamma$, $\Sigma$ and $\Omega$, and that the geometrical properties of $\TC$, $\TS$ and $\TV$ converge to the exact set of normals, curvatures, lengths, areas and volumes associated with the input geometry as $\bar{h}\left(\mathbf{x}\right)\rightarrow 0$.

\subsection{Refinement Loop}

\medskip

\begin{algorithm}[t]
\centering
\caption{Three-dimensional Restricted Delaunay-refinement}

\label{algorithm_restricted_delaunay_refinement}

\begin{minipage}[c]{.900\textwidth}

\small
\smallskip

\begin{algorithmic}[1]
\Function{DelaunayMesh}{$\Gamma,\Sigma,\Omega,\bar{\rho},\bar{\epsilon}(\mathbf{x}),\bar{h}\left(\mathbf{x}\right),\TC,\TS,\TV$}

 \smallskip

 \State \parbox[t]{.900\textwidth}{Form an initial pointwise sampling $X$ such that $X$ is \textit{well-distributed} on $\Gamma$ and $\Sigma$. Compute the Delaunay tessellation $\Del{X}$ and the restricted curve, surface and volume tessellations $\DelC{X}$, $\DelS{X}$ and $\DelV{X}$.\strut}

 \smallskip

 \State \parbox[t]{.900\textwidth}{If some 1-simplex $e\in\DelC{X}$ violates \textsc{BadSimplex1}$(e)$, form the Steiner point $\mathbf{c}_{e}$ associated with $e$, insert $\mathbf{c}_{e}$ into $X$, update $\Del{X}$ and the restricted tessellations $\DelC{X}$, $\DelS{X}$ and $\DelV{X}$ and go to step 3.\strut}

 \smallskip

 \State \parbox[t]{.900\textwidth}{For all vertices $p\in\DelC{X}$ compute $\mathbf{c}_{p}\gets$\textsc{TopoDisk1}$(p)$. If $\mathbf{c}_{p}$ is non-null, insert $\mathbf{c}_{p}$ into $X$, update $\Del{X}$ and the restricted tessellations $\DelC{X}$, $\DelS{X}$ and $\DelV{X}$ and go to step 3.\strut}

 \smallskip

 \State \parbox[t]{.900\textwidth}{If some 2-simplex $f\in\DelS{X}$ violates \textsc{BadSimplex2}$(f)$, form the Steiner point $\mathbf{c}_{f}$ associated with $f$: 
\begin{enumerate}[(a)]\itemsep 0pt
\item If the point $\mathbf{c}_{f}$ lies within a surface ball $B(\mathbf{c}_{e},r)$ associated with some 1-face $e\in\DelC{X}$, insert $\mathbf{c}_{e}$ into $X$ instead, update $\Del{X}$ and the restricted tessellations $\DelC{X}$, $\DelS{X}$ and $\DelV{X}$ and go to step 3.

\item Insert $\mathbf{c}_{f}$ into $X$. If $\mathbf{c}_{f}$ changes the topology of $\DelC{X}$, find the largest adjacent surface ball $B(\mathbf{c}_{e},r)$, delete $\mathbf{c}_{f}$ from $X$ and insert $\mathbf{c}_{e}$ into $X$. update $\Del{X}$ and the restricted tessellations $\DelC{X}$, $\DelS{X}$ and $\DelV{X}$ and go to step 3.

\item Go to step 5.
\end{enumerate}}

%\vspace*{-12pt}

 \smallskip

 \State \parbox[t]{.900\textwidth}{For all vertices $p\in\DelS{X}$ compute $\mathbf{c}_{p}\gets$\textsc{TopoDisk2}$(p)$. If $\mathbf{c}_{p}$ is non-null, insert $\mathbf{c}_{p}$ into $X$, update $\Del{X}$ and the restricted tessellations $\DelC{X}$, $\DelS{X}$ and $\DelV{X}$ and go to step 3.\strut}

 \smallskip

 \State \parbox[t]{.900\textwidth}{If some 3-simplex $\tau\in\DelV{X}$ violates \textsc{BadSimplex3}$(f\tau)$, form the Steiner point $\mathbf{c}_{\tau}$ associated with $\tau$: 
\begin{enumerate}[(a)]\itemsep 0pt
\item If the point $\mathbf{c}_{\tau}$ lies within a surface ball $B(\mathbf{c}_{e},r)$ associated with some 1-face $e\in\DelC{X}$, insert $\mathbf{c}_{e}$ into $X$ instead, update $\Del{X}$ and the restricted tessellations $\DelC{X}$, $\DelS{X}$ and $\DelV{X}$ and go to step 3.

\item If the point $\mathbf{c}_{\tau}$ lies within a surface ball $B(\mathbf{c}_{f},r)$ associated with some 2-face $f\in\DelS{X}$, insert $\mathbf{c}_{f}$ into $X$ instead, update $\Del{X}$ and the restricted tessellations $\DelC{X}$, $\DelS{X}$ and $\DelV{X}$ and go to step 3.

\item Insert $\mathbf{c}_{\tau}$ into $X$. If $\mathbf{c}_{\tau}$ changes the topology of $\DelC{X}$, find the largest adjacent surface ball $B(\mathbf{c}_{e},r)$, delete $\mathbf{c}_{\tau}$ from $X$ and insert $\mathbf{c}_{e}$ into $X$. update $\Del{X}$ and the restricted tessellations $\DelC{X}$, $\DelS{X}$ and $\DelV{X}$ and go to step 3.

\item Insert $\mathbf{c}_{\tau}$ into $X$. If $\mathbf{c}_{\tau}$ changes the topology of $\DelS{X}$, find the largest adjacent surface ball $B(\mathbf{c}_{f},r)$, delete $\mathbf{c}_{\tau}$ from $X$ and insert $\mathbf{c}_{f}$ into $X$. update $\Del{X}$ and the restricted tessellations $\DelC{X}$, $\DelS{X}$ and $\DelV{X}$ and go to step 3.

\item Go to step 7.
\end{enumerate}}

%\vspace*{-12pt}

 \smallskip

 \State \parbox[t]{.900\textwidth}{Return the final restricted Delaunay curve, surface and volume tessellations $\DelC{X}$, $\DelS{X}$ and $\DelV{X}$.\strut} 

 \smallskip

\EndFunction
\end{algorithmic}

\smallskip

\end{minipage}
\end{algorithm}

The Delaunay-refinement algorithm begins by pre-processing the input geometry -- seeking to identify any \textit{sharp-features} inscribed on the input curve and surface collections. These 0- and 1-dimensional features can be induced by both geometrical and topological constraints, including: (i) features that form sharp \textit{creases} or \textit{corners} in $\Gamma$ and/or $\Sigma$, and (ii) features at the apex of non-manifold topological connections. After pre-processing, an initial point-wise sampling of the input curve and surface segments $\Gamma$ and $\Sigma$ is created. Exploiting the discrete representations available, the initial sampling is obtained in this study as a \textit{well-distributed}\footnote{A subset of \textit{seed} vertices $Z$ are sampled from $\mathcal{P}$, such that $Z$ is \textit{well-separated}. Specifically, each new point $z_{n}$ is chosen to maximise the minimum distance to the existing points $z_{1}\dots z_{n-1}$. In this study $n=8$ is used throughout. See \cite{ChengDeyShewchuk} for a provably-good initialisation technique.} subset of the existing vertices $Y\in\mathcal{P}$, where $\mathcal{P}$ is the polyhedral representation of the curve and surface segments $\Gamma$ and $\Sigma$. In the next step, the initial triangulation objects are formed. In this work, the full-dimensional Delaunay tessellation, $\Del{X}$, is built using an incremental Delaunay triangulation algorithm, based on the Bowyer-Watson technique \cite{bowyer81algorithm}. The restricted curve, surface and volumetric triangulations, $\DelC{X}$, $\DelS{X}$ and $\DelV{X}$, are derived from the topology of $\Del{X}$ by explicitly testing for intersections between the faces of the associated Voronoi complex $\Vor{X}$ and the input curve and surface segments $\Gamma$ and $\Sigma$. These queries are computed efficiently by storing the polyhedral geometry $\mathcal{P}$ in an \textsc{aabb}-tree \cite{engwirda2014locally,Alliez09aabbtree}.

\begin{algorithm}[t]
\centering
\caption{Topological Disks \& Termination Criteria}

\label{algorithm_topological_disks}

\begin{minipage}[c]{.475\textwidth}
\centering

\vspace{0pt}
\begin{minipage}{.90\textwidth}

\small
\smallskip

\begin{algorithmic}[1]

\Function{TopoDisk1}{$p$}%\Comment{\{topological $1$-disk about $\mathbf{p}$\}}

 \smallskip

 \State \parbox[t]{.85\textwidth}{Find the set of 1-simplexes $E_{p}\in\DelC{X}$ adjacent to the vertex $p$.\strut}
 
 \smallskip
 
 \State \parbox[t]{.85\textwidth}{If $E_{p}$ is either empty or a valid topological $1$-disk, return \texttt{NULL}. Otherwise, find the 1-simplex $e\in E_{p}$ that maximises the size of the associated surface Delaunay ball $B(\mathbf{c}_{e},r)$ and return $\mathbf{c}_{e}$.\strut}

 \smallskip

\EndFunction
\end{algorithmic}

%\smallskip

\begin{algorithmic}[1]
\Function{TopoDisk2}{$p$}%\Comment{\{topological $2$-disk about $\mathbf{p}$\}}

 \smallskip

 \State \parbox[t]{.85\textwidth}{Find the set of 2-simplexes $F_{p}\in\DelS{X}$ adjacent to the vertex $p$.\strut}
 
 \smallskip
 
 \State \parbox[t]{.85\textwidth}{If $F_{p}$ is either empty or a valid topological $2$-disk, return \texttt{NULL}. Otherwise, find the 2-simplex $f\in F_{p}$ that maximises the size of the associated surface Delaunay ball $B(\mathbf{c}_{f},r)$ and return $\mathbf{c}_{f}$.\strut}

 \smallskip

\EndFunction
\end{algorithmic}

\smallskip

\end{minipage}

\end{minipage}
\vrule{}
\begin{minipage}[c]{.475\textwidth}
\centering

\vspace{0pt}
\begin{minipage}{.90\textwidth}

\small
\smallskip

\begin{algorithmic}[1]
\Function{BadSimplex1}{$e$}%\Comment{\{termination criteria for $1$-simplexes\}}
 
 \smallskip
 
 \State
 \begin{tabular}{ll}
  if $(\epsilon_{1}(e)>\bar{\epsilon}(\mathbf{x}_{e}))$ &\Return \texttt{TRUE} \\
  if $(h\,(e)>\bar{h}(\mathbf{x}_{e}))$                 &\Return \texttt{TRUE} \\
  else, \Return \texttt{FALSE} &
 \end{tabular}

 \smallskip

\EndFunction
\end{algorithmic}

%\smallskip

\begin{algorithmic}[1]
\Function{BadSimplex2}{$f$}%\Comment{\{termination criteria for $2$-simplexes\}}

 \smallskip
 
 \State
 \begin{tabular}{ll}
  if $(\epsilon_{2}(f)>\bar{\epsilon}(\mathbf{x}_{f}))$ &\Return \texttt{TRUE} \\
  if $(h\,(f)>\bar{h}(\mathbf{x}_{f}))$                 &\Return \texttt{TRUE} \\
  if $(\rho_{2}(f)>\bar{\rho})$                         &\Return \texttt{TRUE} \\
  else, \Return \texttt{FALSE} &
 \end{tabular}

 \smallskip

\EndFunction
\end{algorithmic}

%\smallskip

\begin{algorithmic}[1]
\Function{BadSimplex3}{$\tau$}%\Comment{\{termination criteria for $3$-simplexes\}}

 \smallskip
 
 \State
 \begin{tabular}{ll}
  if $(h(\tau)>\bar{h}(\mathbf{x}_{\tau}))$             &\Return \texttt{TRUE} \\
  if $(\rho_{3}(\tau)>\bar{\rho})$                      &\Return \texttt{TRUE} \\
  else, \Return \texttt{FALSE} &
 \end{tabular}

 \smallskip

\EndFunction
\end{algorithmic}

\smallskip

\end{minipage}

\end{minipage}

\end{algorithm}

The main loop of the algorithm proceeds to incrementally refine any restricted 1-, 2- or 3-simplexes found to be in violation of one or more geometrical or topological constraints. Specifically, in steps 3, 5 and 7, any simplexes $e\in\DelC{X}$, $f\in\DelS{X}$ or $\tau\in\DelV{X}$ found to violate a set of local topological, radius-edge, mesh-size or surface-error constraints are refined -- through the introduction of a new Steiner point $\mathbf{c}_{e}$, $\mathbf{c}_{f}$ or $\mathbf{c}_{\tau}$ located at the centre of the associated surface balls $B(\mathbf{c}_{e},r)$, $B(\mathbf{c}_{f},r)$ or circumscribing ball $B(\mathbf{c}_{\tau},r)$. Importantly, refinement proceeds in a hierarchical manner, with the insertion of $\mathbf{c}_{f}$ and $\mathbf{c}_{\tau}$ dependent on several additional constraints designed to preserve the consistency of the lower dimensional tessellations $\DelC{X}$ and $\DelS{X}$. Specifically, in steps 5a--5b and 7a--7d a set of \textit{encroachment} conditions are enforced, with refinement cascading onto lower-dimensional simplexes if a set of geometrical or topological constraints are violated. In steps 5a and 7a, 7c, if $\mathbf{c}_{f}$ or $\mathbf{c}_{\tau}$ are found to lie within the surface balls $B(\mathbf{c}_{e},r)$ or $B(\mathbf{c}_{f},r)$ of an existing curve or surface facet, that facet is instead refined, through the insertion of a Steiner vertex located at the centre of the associated surface ball $\mathbf{c}_{e}$ or $\mathbf{c}_{f}$. This process can be seen simply as an extension of the standard edge-encroachment scheme used in Ruppert's two-dimensional refinement algorithm. In steps 5b and 7b, 7d, if the insertion of $\mathbf{c}_{f}$ or $\mathbf{c}_{\tau}$ is found to modify the restricted triangulations $\DelC{X}$ or $\DelS{X}$, the insertion is deferred onto an adjacent curve or surface facet. Specifically, the point $\mathbf{c}_{f}$ or $\mathbf{c}_{\tau}$ is deleted from $\Del{X}$ and a new Steiner vertex $\mathbf{c}_{e}$ or $\mathbf{c}_{f}$, corresponding to the centre of the largest adjacent surface ball, is inserted instead. This process ensures that the consistency of the curve and surface tessellations $\DelC{X}$ and $\DelS{X}$ is preserved by subsequent refinement operations.

In additional to this element-by-element refinement, the topological consistency of the restricted curve and surface tessellations is also enforced in aggregate, by ensuring that the set of 1- and 2-simplexes $E_{p}\in\DelC{X}$ and $F_{p}\in\DelS{X}$ adjacent to each vertex $p\in\Del{X}$ form locally 1- and 2-manifold features, known as \textit{topological 1- and 2-disks}. Vertices adjacent to non-manifold connections trigger additional refinement operations, with the centres $\mathbf{c}_{e}$ and $\mathbf{c}_{f}$ of the largest adjacent surface balls $B(\mathbf{c}_{e},r)$ and $B(\mathbf{c}_{f},r)$ associated with the simplexes $e\in E_{p}$ and $f\in F_{p}$ inserted as new Steiner vertices until local topological consistency is recovered. Specifically, a cascade of new Steiner vertices are inserted until the topology of $\DelC{X}$ and $\DelS{X}$, sampled at all points $p\in\Del{X}$, is equivalent to that of the input curve and surface complexes $\Gamma$ and $\Sigma$.

The refinement process continues until all radius-edge, mesh-size, surface-error and topological constraints are satisfied for all simplexes $e\in\DelC{X}$, $f\in\DelS{X}$ and $\tau\in\DelV{X}$. The refinement process is priority scheduled, with triangles $f\in\DelS{X}$ and tetrahedrons $\tau\in\DelV{X}$ ordered according to their radius-edge ratios $\rho\left(f\right)$ and $\rho\left(\tau\right)$, ensuring that the element with the \textit{worst} ratio is refined at each iteration. Segments in $\DelC{X}$ are ordered according to the size of their surface balls. Mesh-size constraints are applied with respect to the size of the circumscribing balls associated with each element. Specifically, the mean element sizes $h(e) = 2\,r_{e}$, $h(f) = \sqrt{3}\,r_{f}$ and $h(\tau) = \sqrt{\nicefrac{8}{3}}\,r_{\tau}$ are used throughout, where $h(e)$, $h(f)$ and $h(\tau)$ denote the size associated with segments, triangles and tetrahedrons respectively. The scalar coefficients represent mappings between circumball radii and edge length for equilateral elements. In this study, mesh-size constraints are implemented as $h(e)\leq\alpha \bar{h}(\mathbf{x}_{e})$, $h(f)\leq\alpha \bar{h}(\mathbf{x}_{f})$ and $h(\tau)\leq\alpha \bar{h}(\mathbf{x}_{\tau})$, where $\alpha = \nicefrac{4}{3}$ is a constant factor designed to ensure that mean element size does not, on average, undershoot the target size. The local mesh-size values $\bar{h}(\mathbf{x}_{e})$, $\bar{h}(\mathbf{x}_{f})$ and $\bar{h}(\mathbf{x}_{\tau})$ are evaluated at the centres of the associated circumballs.

\section{Feature Conforming Off-centre Steiner Points}
\label{section_offcentres}

Frontal-Delaunay algorithms are a hybridisation of advancing-front and Delaunay-refinement techniques, in which a Delaunay triangulation is used to define the topology of a mesh while Steiner vertices are inserted consistent with advancing-front type methodologies. In practice, such techniques have been observed to produce very high-quality meshes, inheriting the smooth, semi-structured vertex placement of pure advancing-front methods and the optimal mesh topology and robustness of Delaunay-based approaches. Such techniques have been employed in a number of studies, including, for example \cite{Ungor09OptSteiner,
Rebay93FrontalDelaunay,Mavriplis95FrontalDelaunay,Frey98FrontalDelaunay,Remacle13Quads,
foteinos2010fully,chernikov2012generalized} and previous work by the author \cite{Engwirda2016157,Engwirda2015330,engwirda2014locally}.

%\subsection{Off-centres}

%The off-centre techniques developed in this section exploit the geometrical and topological duality between the Delaunay triangulation and Voronoi complex. Specifically, noting that the cells of the Voronoi complex is a closest-point map for vertices in the Delaunay triangulation, something something something here... 

\subsection{Point-placement Strategy (Edge Segments)}

\medskip

The \textit{off-centre} strategy used to refine curve segments is based on methods previously developed by the author \cite{Engwirda2016157,Engwirda2015330,engwirda2014locally} for surface and volume refinement. Two candidate Steiner vertices are considered. Type~I vertices, $\mathbf{c}^{(1)}$, are equivalent to conventional element circumcentres (positioned at the centre of the associated surface balls), and are used to preserve global convergence. Type~II vertices, $\mathbf{c}^{(2)}$, are so-called \textit{size-optimal} points, and are designed to satisfy mesh-size constraints in a locally optimal fashion. Given a \textit{refinable} 1-simplex $e\in\DelC{X}$, the Type~II vertex $\mathbf{c}^{(2)}$ is positioned at an intersection of the curve network $\Gamma$, and a sphere $S_{\sigma}$ of radius $\bar{h}_{\sigma}$, centred on a vertex $\mathbf{x}_{1}\in e$. The vertex $\mathbf{c}^{(2)}$ is positioned such that it forms an edge candidate $\sigma$ about the \textit{frontal} vertex $\mathbf{x}_{1}$, such that its size $h\left(\sigma\right)$ satisfies $\bar{h}(\mathbf{x})$. Specifically, the length of $\sigma$ is computed from local mesh-size information, such that:
\begin{equation}
{\bar{h}_{\sigma}} = \tfrac{1}{2}\left({\bar{h}(\mathbf{x}_{1})}+{\bar{h}(\mathbf{c}^{(2)})}\right).
\end{equation}
For non-uniform $\bar{h}\left(\mathbf{x}\right)$, this expression is weakly non-linear, and an iterative procedure is used to obtain an approximate solution. In the case of multiple intersections between the curve network $\Gamma$ and the sphere $S_{\sigma}$, the point $\mathbf{c}_{j}^{(2)}$ that minimises the angle to the \textit{frontal} vector $\mathbf{v}$ is chosen, where $\mathbf{v}$ is oriented from the frontal vertex $\mathbf{x}_{1}$ to the centre of the surface ball $B(\mathbf{c}_{e},r)$. See Figure~\ref{figure_restricted_edge}(iii) for additional illustration.

%The positioning of size-optimal Type~II Steiner vertices for curve facets is illustrated in Figure~\ref{figure_restricted_edge}(iii).

\subsection{Point-placement Strategy (Triangles \& Tetrahedrons)}

\medskip

Similar techniques are used to refine triangles and tetrahedrons, with two candidate Steiner vertices used to balance local optimality and global convergence guarantees. These methods are presented by the author in detail in \cite{Engwirda2016157,Engwirda2015330}.

%The point-placement rules used to refine surface and volume elements follow a similar methodology to that described previously for curve segments, with two candidate Steiner vertices used to balance local optimality with guaranteed convergence and provably-good behaviour. These methods are presented in detail in \cite{Engwirda2016157,Engwirda2015330}. 

\subsection{Point-placement Strategy (Off-centre Selection)}

\medskip

Given the sets of Type~I and Type~II off-centres $\mathbf{c}_{e,\,f,\,\tau}^{(1)}$ and $\mathbf{c}_{e,\,f,\,\tau}^{(2)}$ available for curve, surface and volume elements, respectively, the positions of the associated refinement points $\mathbf{c}_{e}$, $\mathbf{c}_{f}$ and $\mathbf{c}_{\tau}$ are calculated. These points are selected to satisfy the \textit{limiting} local constraints, setting
\begin{equation}
\label{eqn_off_centre_selection}
\mathbf{c}_{e}=\left\{
\begin{array}{ll}
\mathbf{c}^{(2)}_{e}, & 
\text{ if }\big(d^{(2)}_{e} \leq d^{(1)}_{e}\big)
\\[1ex]
\mathbf{c}^{(1)}_{e}, &
\text{ otherwise}
\end{array}
\right.
\quad \text{and}\quad
\mathbf{c}_{f,\,\tau}=\left\{
\begin{array}{ll}
\mathbf{c}^{(2)}_{f,\,\tau}, & 
\text{ if }\big(d^{(2)}_{f,\,\tau} \leq d^{(1)}_{f,\,\tau}\big) \text{ and } \big(d^{(2)}_{f,\,\tau} \geq r_{0}\big)
\\[1ex]
\mathbf{c}^{(1)}_{f,\,\tau}, &
\text{ otherwise }
\end{array}
\right.\quad
\end{equation}
where $d^{(i)}=\|\mathbf{c}^{(i)}-\mathbf{c}_{0}\|$ are distances from the centre of the frontal facet to the Type~I and Type~II points, respectively and $r_{0}$ is the radius of the diametric ball associated with the frontal face or vertex. This cascading selection criteria ensures that refinement scheme smoothly degenerates to that of a conventional circumcentre-based Delaunay-refinement strategy in limiting cases, while using locally optimal points where possible. Specifically, these constraints guarantee that the refinement points for curve, surface and volume elements lie within a local \textit{safe} region on the Voronoi complex -- being positioned on an adjacent Voronoi sub-face and bound between the circumcentre of the element itself and the diametric ball of the associated frontal entity. See previous work by the author \cite{Engwirda2016157,Engwirda2015330,engwirda2014locally} for additional details.

\subsection{Refinement Order}
 
\medskip
 
In addition to off-centre point-placement rules, the Frontal-Delaunay algorithm also relies on changes to the order in which elements are refined. To better mimic the behaviour of advancing-front type methods, elements are refined only if they are adjacent to existing \textit{frontal} entities. In the case of curve facets $e\in\DelC{X}$, the frontal vertex $\mathbf{x}_{i}\in e$ must be shared by at least one adjacent facet $e_{j}\in\DelC{X}$ that is \textit{converged} -- satisfying its associated topological and geometrical constraints. In the case of surface triangles $f\in\DelS{X}$ and interior tetrahedrons $\tau\in\DelV{X}$, the frontal facet $e_{0}\in f$, $f_{0}\in \tau$ must either be a converged simplex $e_{j}\in\DelC{X}$, $f_{j}\in\DelS{X}$ or be shared by an adjacent simplex $f_{j}\in\DelS{X}$, $\tau_{j}\in\DelV{X}$ satisfying its associated constraints. In rare cases where no frontal simplex can be found (such as in the initial stages of refinement, where all faces in $\Del{X}$ are still very coarse with respect to $\bar{h}(\mathbf{x})$), standard circumcentre-based refinement is used as a fall-back, ensuring convergence. Use of this type of implicit frontal boundary between converged and un-converged elements is a common feature of Frontal-Delaunay algorithms, with similar approaches used by, for example, \cite{Rebay93FrontalDelaunay,Mavriplis95FrontalDelaunay,Frey98FrontalDelaunay,Remacle13Quads}.

\subsection{Termination, Convergence \& Correctness}

\medskip

%The off-centre refinement rules are derived with respect to properties associated with the underlying Voronoi diagram. Importantly, by forcing new Steiner vertices to lie along sub-faces in $\Vor{X}$, it is guaranteed that the distribution of mesh vertices remains \textit{well-separated} throughout the refinement process. This behaviour ensures that the algorithm does not create arbitrarily short edges, and is guaranteed to converge as a result. For the sake of brevity, a full proof of termination or correctness is not included here, but it is important to note that constraints on element radius-edge ratios $\rho\left(f\right)$, element size $h\,(\mathbf{x})$, surface discretisation error $\epsilon\left(f\right)$ and topological consistency are satisfied \textit{by definition}, provided that termination of the algorithm is achieved in practice. The development of a suitable theoretical model for the new Frontal-Delaunay algorithm is the subject of a forthcoming publication.

The termination and convergence of the Frontal-Delaunay refinement algorithm can be analysed by considering the behaviour of the point-placement rules described previously. For the sake of brevity, a full analysis is not included here, instead a theoretical `sketch' is presented. Firstly, it is important to note that the off-centre vertices selected by the algorithm reduce to standard element circumcentres in limiting cases: 
\begin{remark}
\label{remark_termination_b}
Let $\bar{h}(\boldsymbol{x})\geq\bar{h}_{0}, \bar{h}_{0}\in\mathbb{R}^{+}$ be a positive mesh-size function and $\DelC{X_{k}}$, $\DelS{X_{k}}$ and $\DelV{X_{k}}$ be the restricted triangulation objects after $k$ refinement steps. Given a refinable simplex $e\in\DelC{X_{k}}$, $f\in\DelS{X_{k}}$ or $\tau\in\DelV{X_{k}}$, use of the Type~II point-placement scheme is `declined' if $r \leq \bar{h}_{\sigma}$, where $r$ is the radius of the circumscribing ball $B(\boldsymbol{x}_{e,f,\tau},r)$ and $\bar{h}_{\sigma}$ is the local element `length' computed by the Type~II point-placement rule.
\end{remark}
This behaviour is a consequence of the off-centre selection rules (\ref{eqn_off_centre_selection}), requiring that the refinement point for any 1-, 2- or 3-simplex is the off-centre candidate $\mathbf{c}^{(1)}$ or $\mathbf{c}^{(2)}$ that minimises the distance to the centre of the diametric ball $\operatorname{B}(\mathbf{c}_{0},r_{0})$ associated with the frontal face or vertex. Noting that $\mathbf{c}^{(1)}$ and $\mathbf{c}^{(2)}$ are located on an adjacent segment of $\Vor{X}$, it is clear that $\|\mathbf{c}^{(2)} - \mathbf{c}_{0}\| \leq \|\mathbf{c}^{(1)} - \mathbf{c}_{0}\|$ only if $r \geq \bar{h}_{\sigma}$. 

\medskip

Considering that standard, circumcentre-type insertion rules are recovered when elements become sufficiently small, the overall termination of the algorithm is equivalent to that of a conventional circumcentre-based scheme. In \cite{Rineau2008}, Rineau and Yvinec analyse such an approach, and have shown that, under the assumption of non-acute input, termination is guaranteed provided that:
\begin{equation}
\bar{\rho}_{f} \geq \Big(\!\sqrt{2} + 2\Big)\nu_{0} 
\quad \text{and}\quad
\bar{\rho}_{\tau} \geq \Big(\!\sqrt{2} + 2\Big)\nu_{0}\Big(\nu_{0}+2\Big)
\end{equation} 
where $\nu_{0} = \nicefrac{2\mu_{0}}{\gamma_{0}}$ is a mesh-size ratio, given $\mu_{0}$ as the maximum of $\bar{h}(\mathbf{x})$ over the surface patches $\Sigma$ and $\gamma_{0}$ as the minimum of $\bar{h}(\mathbf{x})$ over the volumes $\Omega$. Typically, the algorithm is found to outperform these bounds in practice.

\medskip

Finite termination leads directly to a number of useful auxiliary guarantees on both the nature and quality of the output tessellation. Adopting the conventional terminology, meshes generated using the Delaunay-refinement and Frontal-Delaunay refinement algorithms presented here can be considered to be \textit{provably-good}:
\begin{remark}
If the meshing algorithm terminates after $k$ steps, the restricted curve, surface and volume sub-complexes $\DelC{X_{k}}$, $\DelS{X_{k}}$ and $\DelV{X_{k}}$ satisfy the following properties:

\smallskip

{\normalfont \textbf{(A) size \& shape-quality:}} The size and shape of elements in the output mesh satisfy several constraints. Specifically, all elements $e\in\DelC{X_{k}}$, $f\in\DelS{X_{k}}$ and $\tau\in\DelV{X_{k}}$ contain bounded radius-edge ratios and circumball sizes, such that: (i) $\rho(f) \leq \bar{\rho}_{f}$ and $\rho(\tau) \leq \bar{\rho}_{\tau}$, and (ii) $h(e) \leq \alpha \bar{h}(\mathbf{x}_{e})$, $h(f) \leq \alpha \bar{h}(\mathbf{x}_{f})$ and $h(\tau) \leq \alpha \bar{h}(\mathbf{x}_{\tau})$, where $\alpha = \nicefrac{4}{3}$ and $h(e) = 2\,r_{e}$, $h(f) = \sqrt{3}\,r_{f}$ and $h(\tau) = \sqrt{\nicefrac{8}{3}}\,r_{\tau}$ are scaled circumball radii.

\smallskip

{\normalfont \textbf{(B) approximation error:}} The surface discretisation error associated with the output mesh is bounded by a one-sided threshold on the Hausdorff distance. Specifically, all elements $e\in\DelC{X_{k}}$ and $f\in\DelS{X_{k}}$, contain bounded surface error measures, such that $\epsilon(e) \leq \bar{\epsilon}(\mathbf{x}_{e})$ and $\epsilon(f) \leq \bar{\epsilon}(\mathbf{x}_{f})$.

\smallskip

{\normalfont \textbf{(C) topological consistency:}} The output mesh is consistent with the topology of the input geometry, consisting of curve and surface tessellations that are locally 1- and 2-manifold in their interiors, and share the degree of the input features at their boundaries.
\end{remark}
The algorithm maintains a queue of \textit{bad} elements, $e\in\DelC{X_{k}}$, $f\in\DelS{X_{k}}$, $\tau\in\DelV{X_{k}}$ and vertices $p \in \Del{X}$, in violation of one or more local constraints. Given that termination of the algorithm is guaranteed, it is clear that the refinement queues must become empty, and that the output mesh satisfies the requisite geometrical and topological constraints as a result.

\medskip

The development of a full theoretical model of termination, convergence and correctness for the new algorithms presented here is the subject of a forthcoming publication.

\section{Protecting Sharp Angles Between Curves}
\label{section_protecting_angles}

Like all Delaunay-refinement type methods, the restricted Delaunay-refinement and Frontal-Delaunay algorithms presented here suffer from issues of non-convergence when the input domain contains sharply acute features. In such cases, a special-case pre-processing phase is required to identify and to \textit{protect} such features. In this study, a technique to protect sharp angles subtended by 1-dimensional features is described. The development of a generalised procedure for sharp features between 2-dimensional surface patches is deferred for future investigation. Specifically, a process modelled on the two-dimensional \textit{corner-lopping} procedure of Pav and Walkington \cite{pav2005delaunay} and the three-dimensional \textit{protective-collar} techniques of Rand and Walkington \cite{rand20083d,rand2009collars} is adopted, in which a small subset of the domain, adjacent to any sufficiently sharp features, is pre-processed and \textit{quarantined} from subsequent refinement operations. Note that this collar-based approach is fundamentally different from the standard \textit{protecting-ball} type techniques utilised in other (statically weighted) restricted Delaunay-refinement algorithms \cite{cheng2008practical,ChengDeyShewchuk,jamin2013cgalmesh}. Specifically, in the current work, a standard (unweighted) Delaunay triangulation is forced to conform to sharp-features in the input geometry via a judicious arrangement of local Steiner vertices.

Given a sharply acute internal angle $\mathcal{A}_{ij} \leq \nicefrac{\pi}{3}$ formed by any pair of segments $i,j$ in the curve network $\Gamma$, the protection process proceeds in a staged fashion: (i) a vertex $\mathbf{x}_{\mathcal{A}}$ is introduced at the apex of the sharp angle $\mathcal{A}_{ij}$, (ii) two new vertices $\mathbf{x}_{i},\mathbf{x}_{j}$ are positioned along the incident curve segments $i,j\in\Gamma$, such that an isosceles triangle candidate $[\mathbf{x}_{\mathcal{A}},\mathbf{x}_{i},\mathbf{x}_{j}]$ is formed. Specifically, the points $\mathbf{x}_{i},\mathbf{x}_{j}$ are positioned at the intersection of a ball of radius $r_{\mathcal{A}}$, centred at $\mathbf{x}_{\mathcal{A}}$, and the curve network $\Gamma$. Ideally, the radius $r_{\mathcal{A}}$ should be chosen to reflect the local-feature-size at the apex of the sharp angle $\operatorname{lfs}(\mathbf{x}_{\mathcal{A}})$. In practice, such a quantity is hard to compute reliably, and the radius $r_{\mathcal{A}}$ is computed using an iterative procedure in the present work instead. In this process, the local mesh-size $\bar{h}(\mathbf{x}_{\mathcal{A}})$ is used as an initial guess for the radius $r_{\mathcal{A}}$. The radius is then iteratively reduced until: (i) there exist exactly two intersections between the ball $\operatorname{B}(\mathbf{x}_{\mathcal{A}}, r_{\mathcal{A}})$ and the curve network $\Gamma$, and (ii) the ball $\operatorname{B}(\mathbf{x}_{\mathcal{A}}, \beta r_{\mathcal{A}})$ is empty of intersections with other balls centred on protected features in $\Gamma$. Here, the scalar $\beta\geq 1$ is a \textit{spacing-factor}, ensuring that adjacent balls are sufficiently well separated. In this study $\beta=\nicefrac{3}{2}$ is used. The protection procedure presented here is similar to the \textsc{SplitBall} operator described in \cite{pav2005delaunay} for two-dimensional piecewise smooth domains.

Noting that such a set of protecting balls is disjoint, and that the candidate vertices describe sets of \textit{well-centred}\footnote{Simplexes possessing circumcentres interior to the hull of the element.} isosceles triangles, it is clear that the Delaunay tessellation $\Del{X}$ contains both the protected edge segments $[\mathbf{x}_{\mathcal{A}},\mathbf{x}_{i}]$ and $[\mathbf{x}_{\mathcal{A}},\mathbf{x}_{j}]$, in addition to the triangles $[\mathbf{x}_{\mathcal{A}},\mathbf{x}_{i},\mathbf{x}_{j}]$, thus constituting a conforming Delaunay triangulation of the sharp features in $\Gamma$. Clearly, these protected elements are also automatically included in the restricted sub-complexes $\DelC{X}$ and $\DelS{X}$. The main loop of the Delaunay-refinement algorithm is modified to ensure that these protected elements are preserved throughout the refinement passes. In this study, a simple topological constraint is enforced: any new Steiner vertex found to delete a protected edge $e_{\mathcal{A}}\in\DelC{X}$ is \textit{rejected}. In practice, this means that a narrow halo of low quality elements adjacent to sharp features are tolerated. The use of topology-based vertex rejection, as opposed to weighted protecting-ball filtering, was found to reduce the size of this halo region in practice. %See Figure~\ref{figure_protecting_curves} for additional information. 

\section{Sliver Suppression}
\label{section_sliver}

Slivers are a class of low-quality tetrahedral elements that occur in three-dimensional Delaunay tessellations. Consisting of four vertices positioned in a thin `kite'-like configuration, sliver elements are typically of very low shape-quality -- possessing pathologically small dihedral angles, but relatively small radius-edge ratios. Sliver elements are not guaranteed to be eliminated by standard Delaunay-based refinement schemes, including the Frontal-Delaunay algorithm presented previously. Various strategies designed to remove sliver elements are known to exist, including non-linear optimisation methods based on \textit{sliver-exudation} \cite{Cheng00SliverExudation} and \textit{topological-optimisation} \cite{Klinger08Stellar}. In this study, a simple method for the suppression of sliver elements is employed, in which slivers are eliminated through additional refinement operations. Following \cite{gosselinvollenref11}, any tetrahedron $\tau_{i}\in\DelV{X}$ with a small \textit{volume-length} ratio $v(\tau_{i})\leq\bar{v}$ is marked for refinement, where $\bar{v}$ is a user-defined lower-bound on element volume-length ratios. Previous studies \cite{gosselinvollenref11,engwirda2014locally} have shown that this modified refinement algorithm is convergent for $\bar{v}\leq\nicefrac{1}{3}$. Noting that the volume-length ratio is a \textit{robust} measure of element quality, known to detect all classes of low-quality tetrahedrons, the resulting meshes are of guaranteed quality, with bounded element dihedral angles and aspect ratios. The Frontal-Delaunay algorithm presented previously was modified to impose additional bounds on element volume-length ratios during the tetrahedral refinement phase. Additional details can be found in \cite{Engwirda2015330,engwirda2014locally}.

\section{Experimental Results}
\label{section_results}

The performance of the Frontal-Delaunay algorithm presented in Sections~\ref{section_restricted_refinement}, \ref{section_offcentres} and \ref{section_protecting_angles} was investigated experimentally, with the method used to mesh a series of benchmark problems. The algorithm was implemented in \cpp and compiled as a 64-bit executable. The Frontal-Delaunay algorithm has been implemented as part of the \texttt{JIGSAW} meshing package, currently available online \cite{JIGSAWpackage} or by request from the author. The Frontal-Delaunay implementation is referred to as \texttt{JGSW-FD} throughout, with the suffix `\texttt{-FD}' denoting the `Frontal-Delaunay' method. In order to provide additional performance information, the well-known \texttt{CGALMESH} implementation \cite{jamin2013cgalmesh} was also used to mesh the same set of benchmark problems. The \texttt{CGALMESH} algorithm was sourced from version 4.6 of the \texttt{CGAL} package \cite{cgal:pt-t3-15a,cgal:pt-tds3-15a} and was compiled as a 64-bit library. The \texttt{CGALMESH} algorithm is referred to as \texttt{CGAL-DR} throughout, with the suffix `\texttt{-DR}' denoting `Delaunay-refinement'. All tests were completed on a Linux platform using a single core of an Intel i7 processor. Visualisation and post-processing was completed using \texttt{MATLAB}.

\begin{figure}[t]
\centering

{
\footnotesize
\tabulinesep=2pt

\smallskip

\begin{tabu} {c|c|c}

\parbox[c][1em][b]{.31\textwidth}{\center (\texttt{JGSW-FD}): $\,|\DelS{X}|=6,842,\, t=0.69\,\mathrm{sec}$}
&
\parbox[c][1em][b]{.31\textwidth}{\center (\texttt{JGSW-FD}): $\,|\DelS{X}|=8,868,\, t=0.66\,\mathrm{sec}$}
&
\parbox[c][1em][b]{.31\textwidth}{\center (\texttt{JGSW-FD}): $|\DelV{X}|=128,825,\, t=3.02\,\mathrm{sec}$}
 
\\

\begin{minipage}[c]{0.31\textwidth}
\begin{center}
\includegraphics[width=5.10cm]{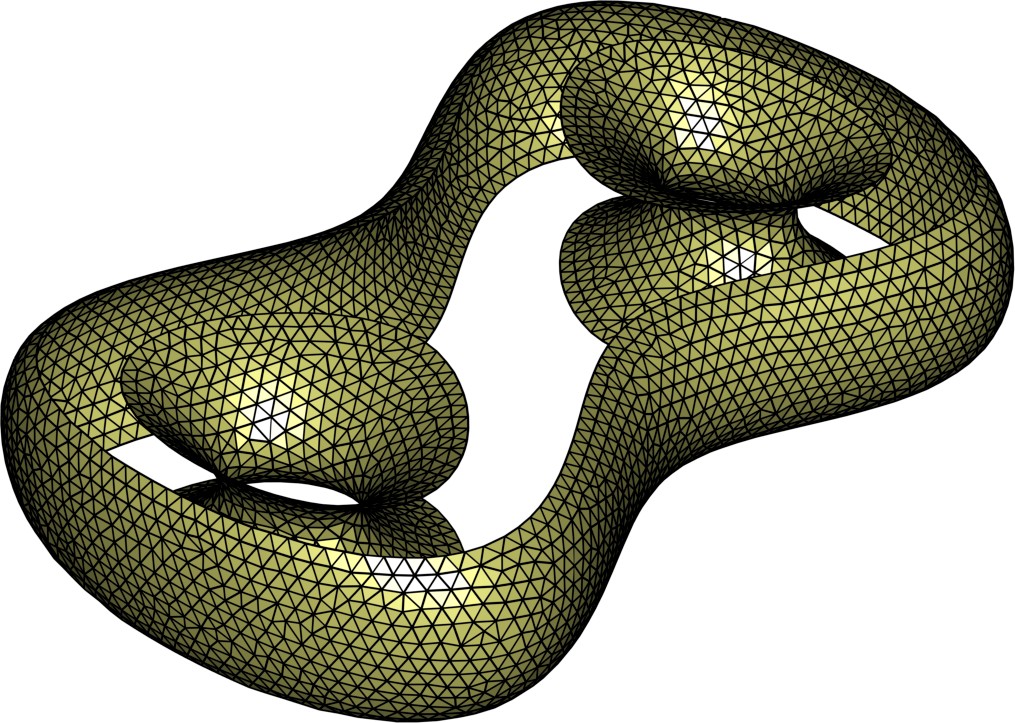} 
\end{center}
\end{minipage} &
\begin{minipage}[c]{0.31\textwidth}
\begin{center}
\includegraphics[width=4.40cm]{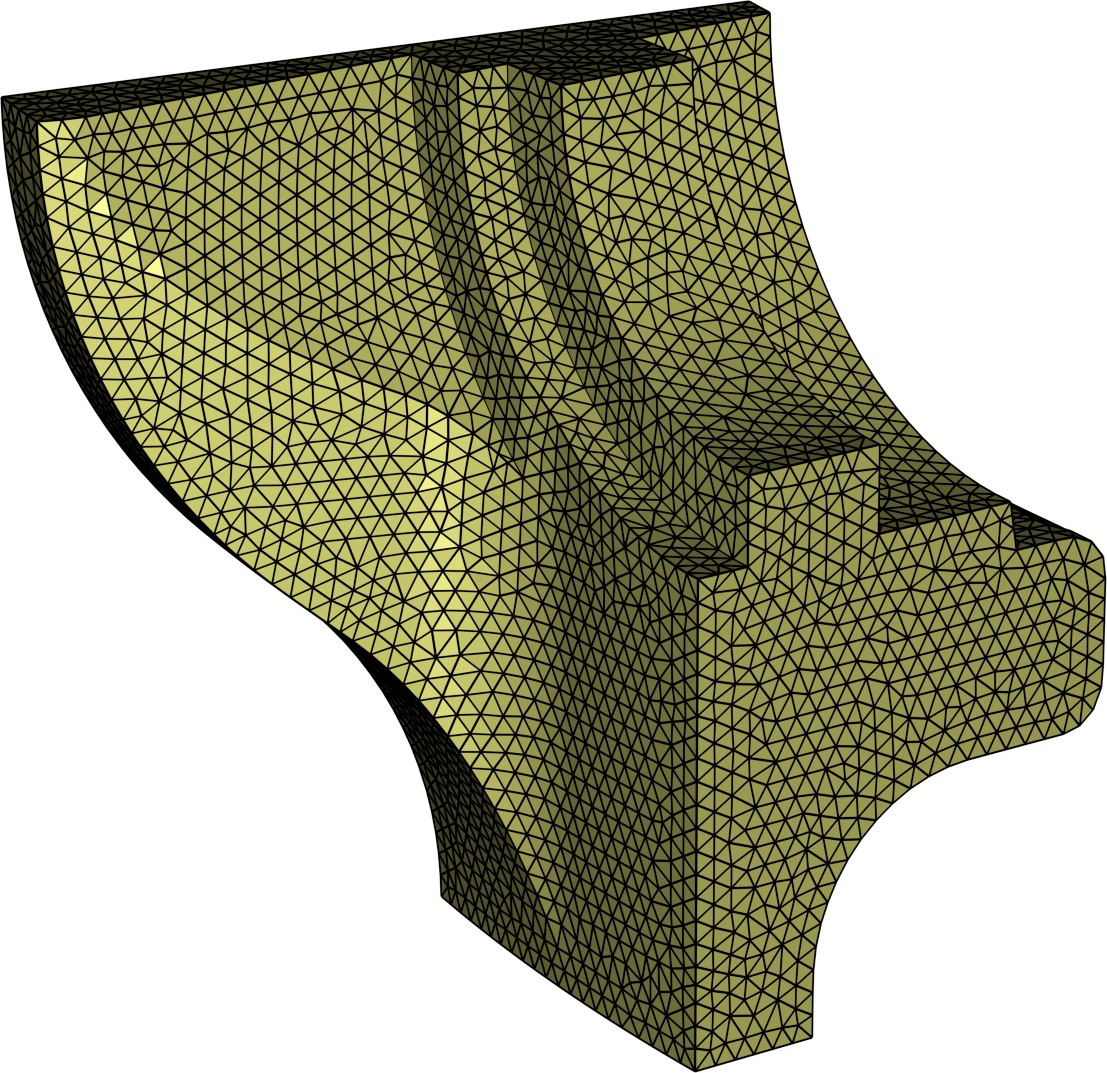} 
\end{center}
\end{minipage} &
\begin{minipage}[c]{0.31\textwidth}
\begin{center}
\includegraphics[width=4.90cm]{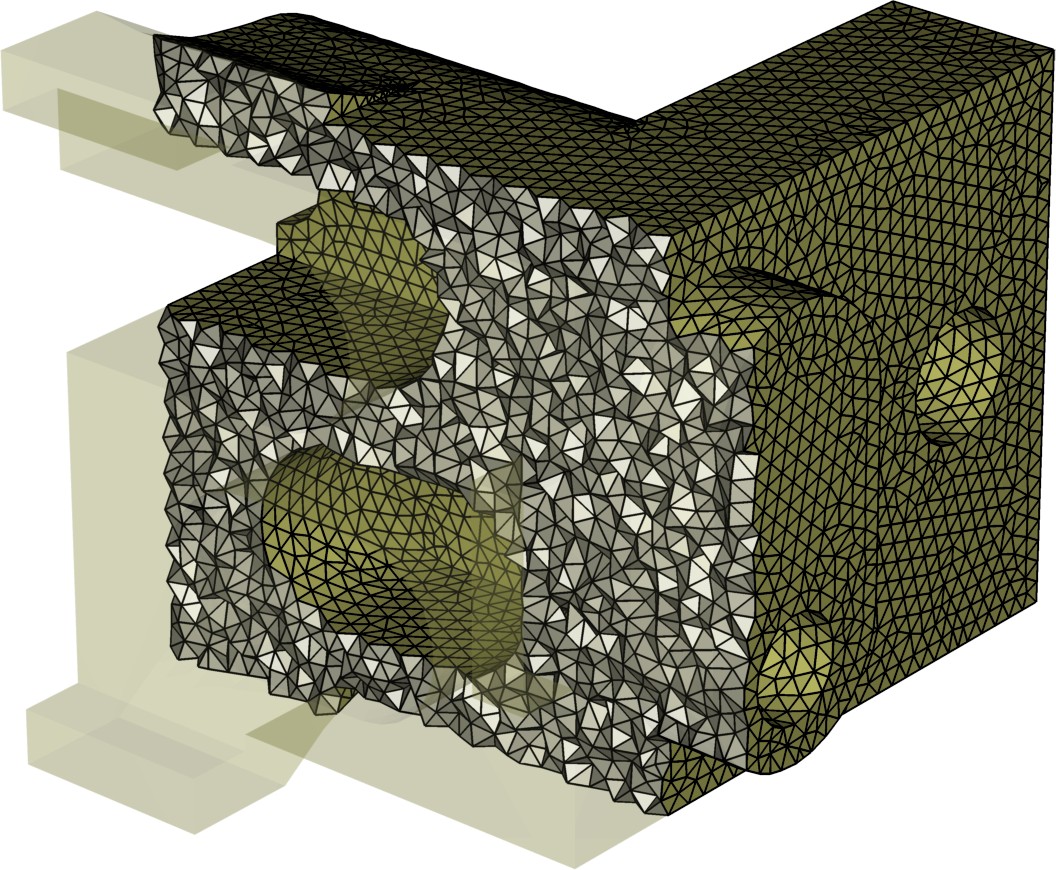} 
\end{center}
\end{minipage}

\\

\begin{minipage}[c]{0.31\textwidth}
\begin{center}
\includegraphics[width=4.90cm]{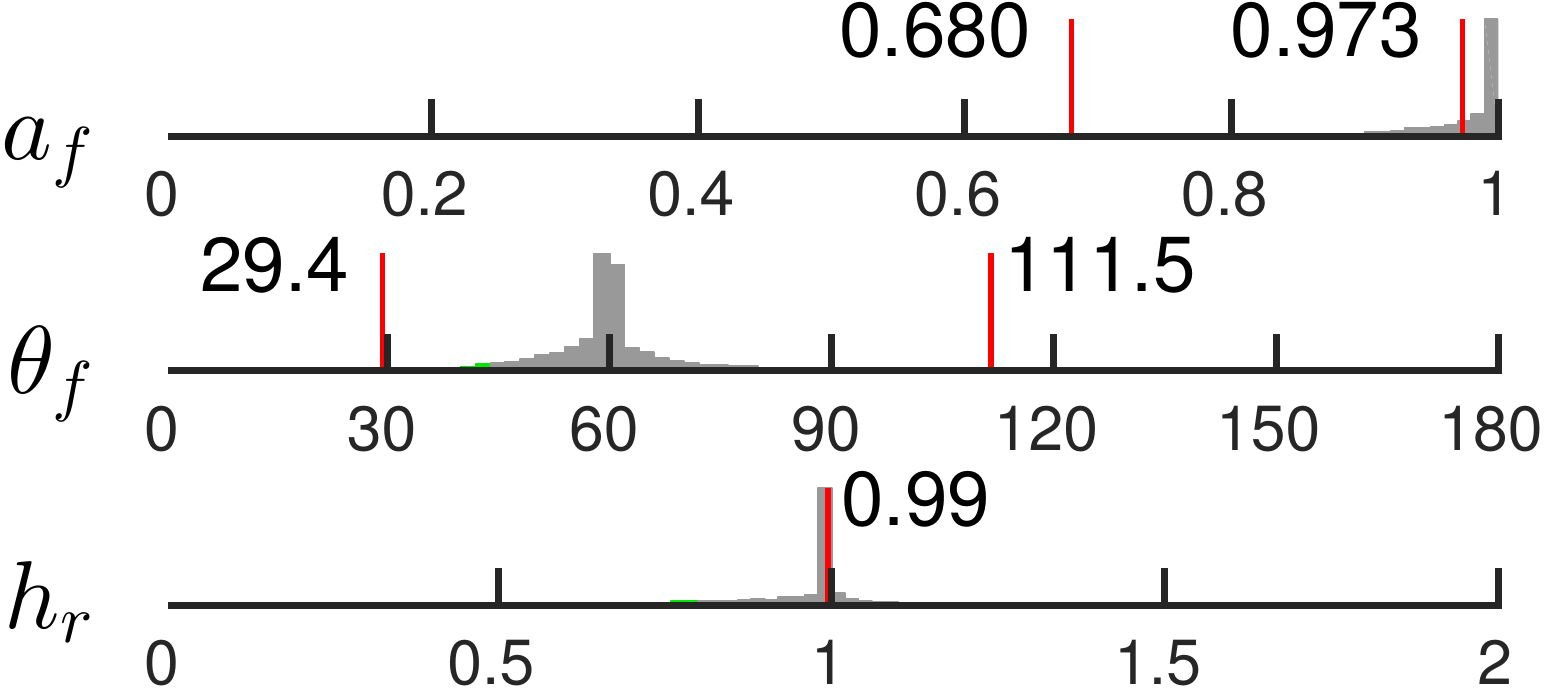} 
\end{center}
\end{minipage} &
\begin{minipage}[c]{0.31\textwidth}
\begin{center}
\includegraphics[width=4.90cm]{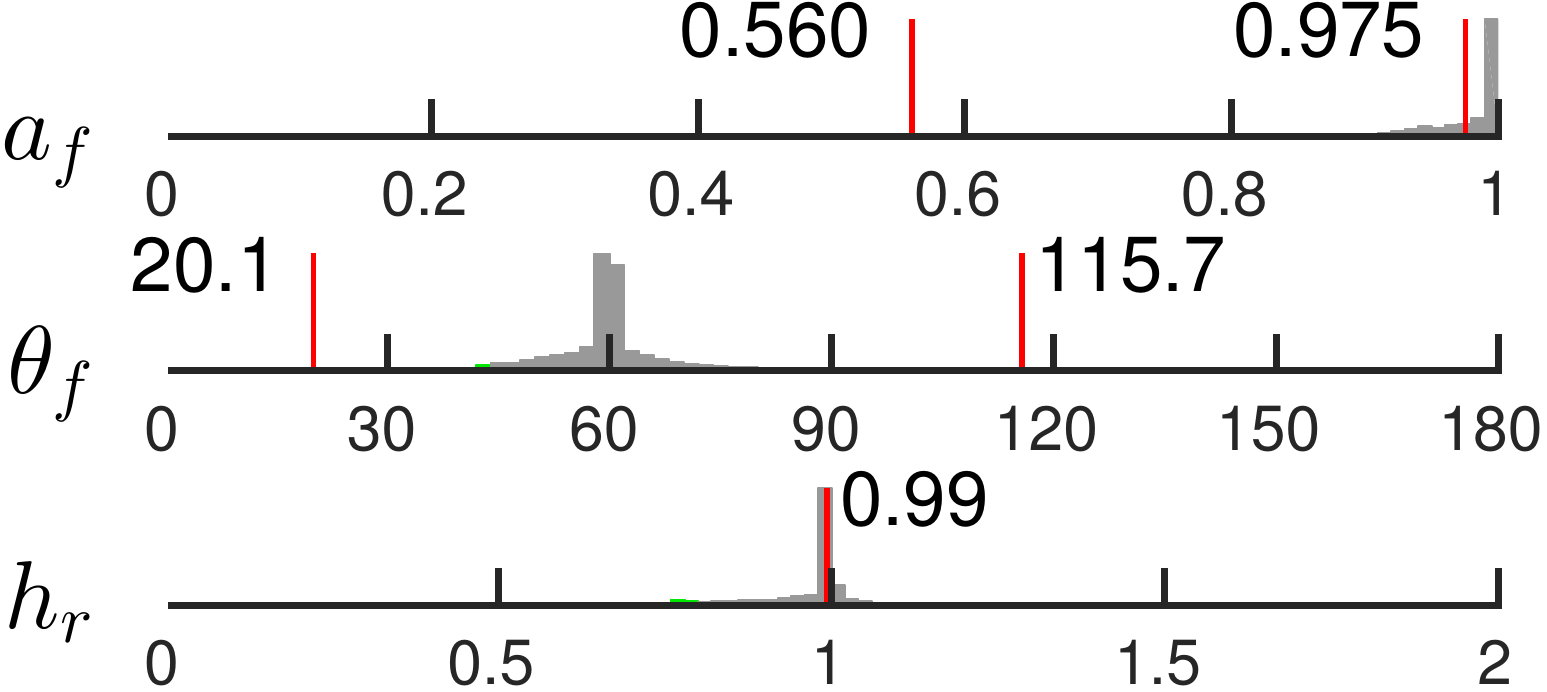} 
\end{center}
\end{minipage} &
\begin{minipage}[c]{0.31\textwidth}
\begin{center}
\includegraphics[width=4.90cm]{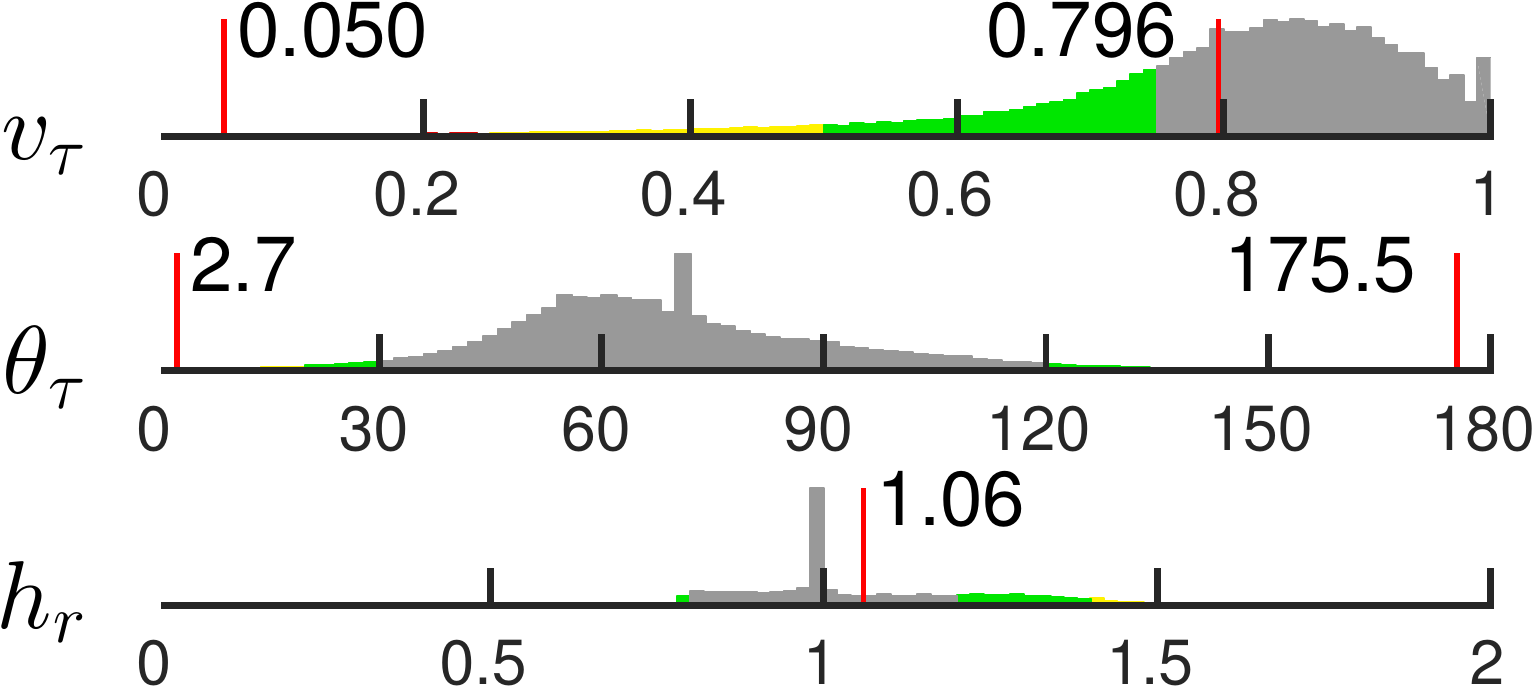} 
\end{center}
\end{minipage}

\\
\hline
\\

\parbox[c][1em][b]{.31\textwidth}{\center (\texttt{CGAL-DR}): $\,|\DelS{X}|=7,208,\, t=0.52\,\mathrm{sec}$}
&
\parbox[c][1em][b]{.31\textwidth}{\center (\texttt{CGAL-DR}): $\,|\DelS{X}|=9,100,\, t=1.45\,\mathrm{sec}$}
&
\parbox[c][1em][b]{.31\textwidth}{\center (\texttt{CGAL-DR}): $|\DelV{X}|=137,899,\, t=2.97\,\mathrm{sec}$}
 
\\

\begin{minipage}[c]{0.31\textwidth}
\begin{center}
\includegraphics[width=5.10cm]{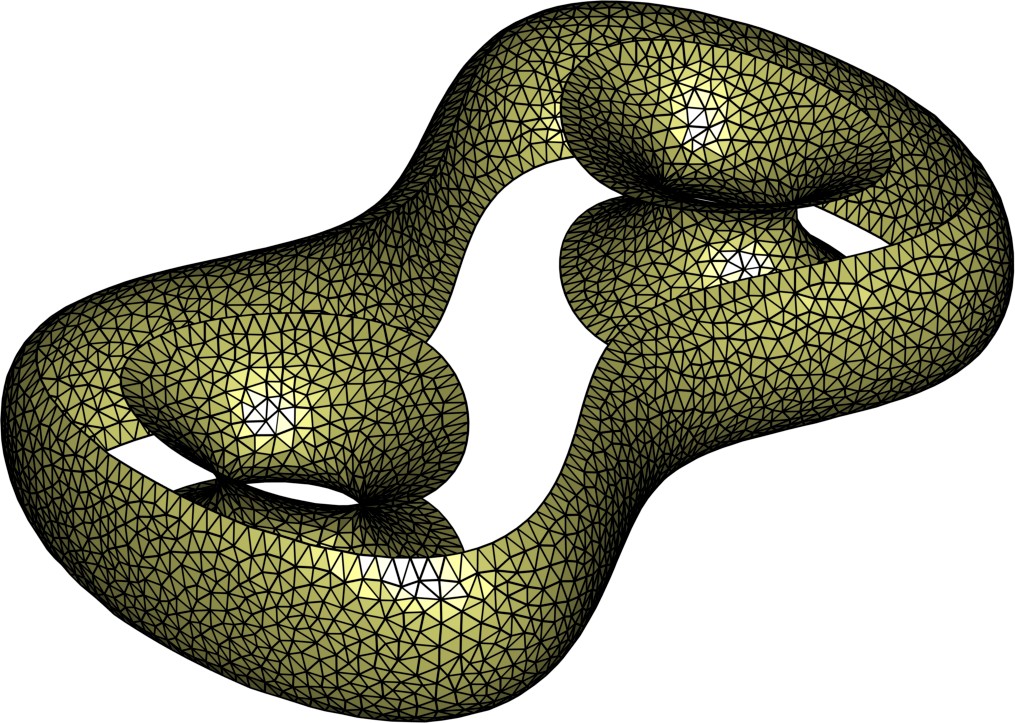} 
\end{center}
\end{minipage} &
\begin{minipage}[c]{0.31\textwidth}
\begin{center}
\includegraphics[width=4.40cm]{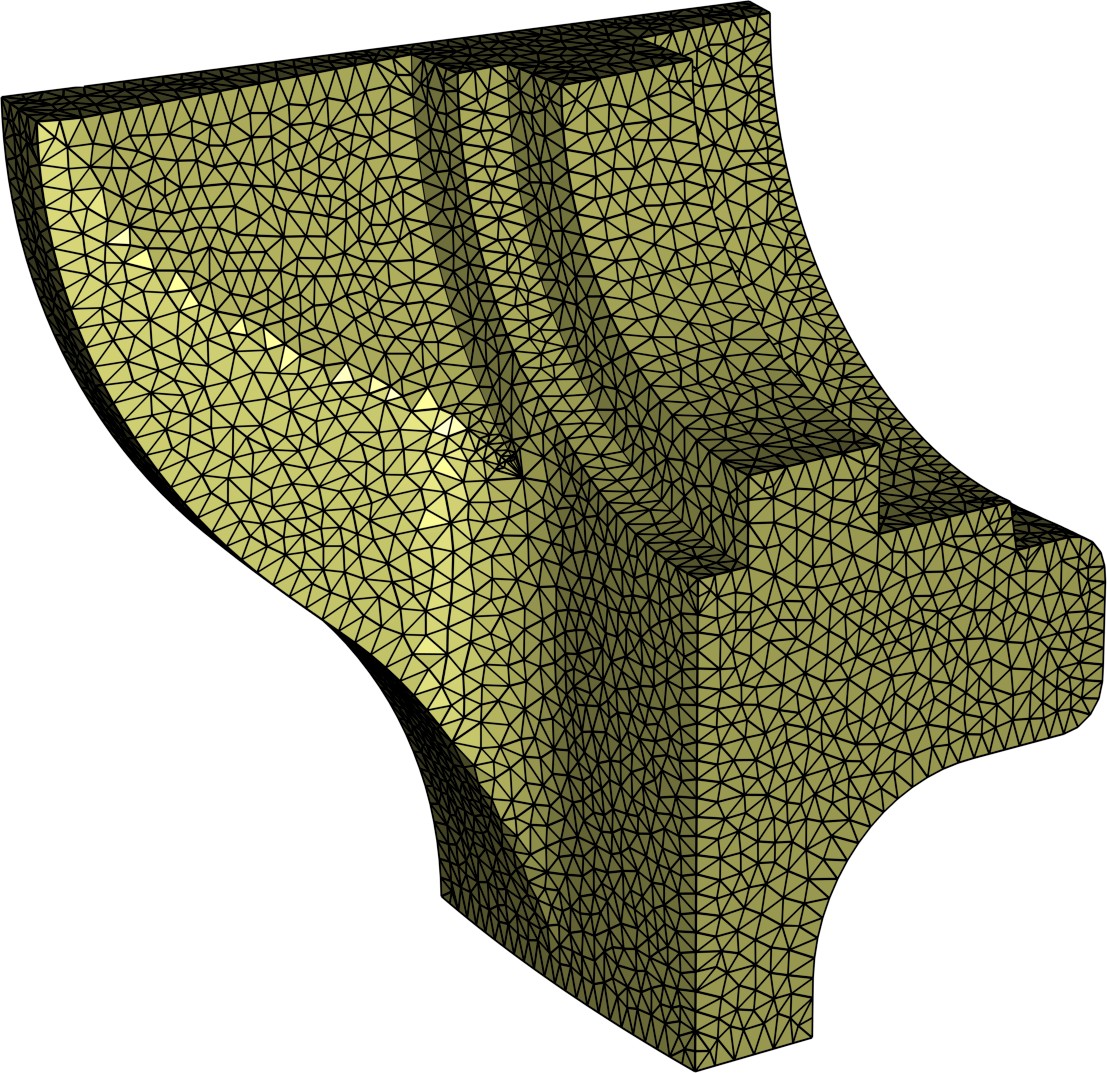} 
\end{center}
\end{minipage} &
\begin{minipage}[c]{0.31\textwidth}
\begin{center}
\includegraphics[width=4.90cm]{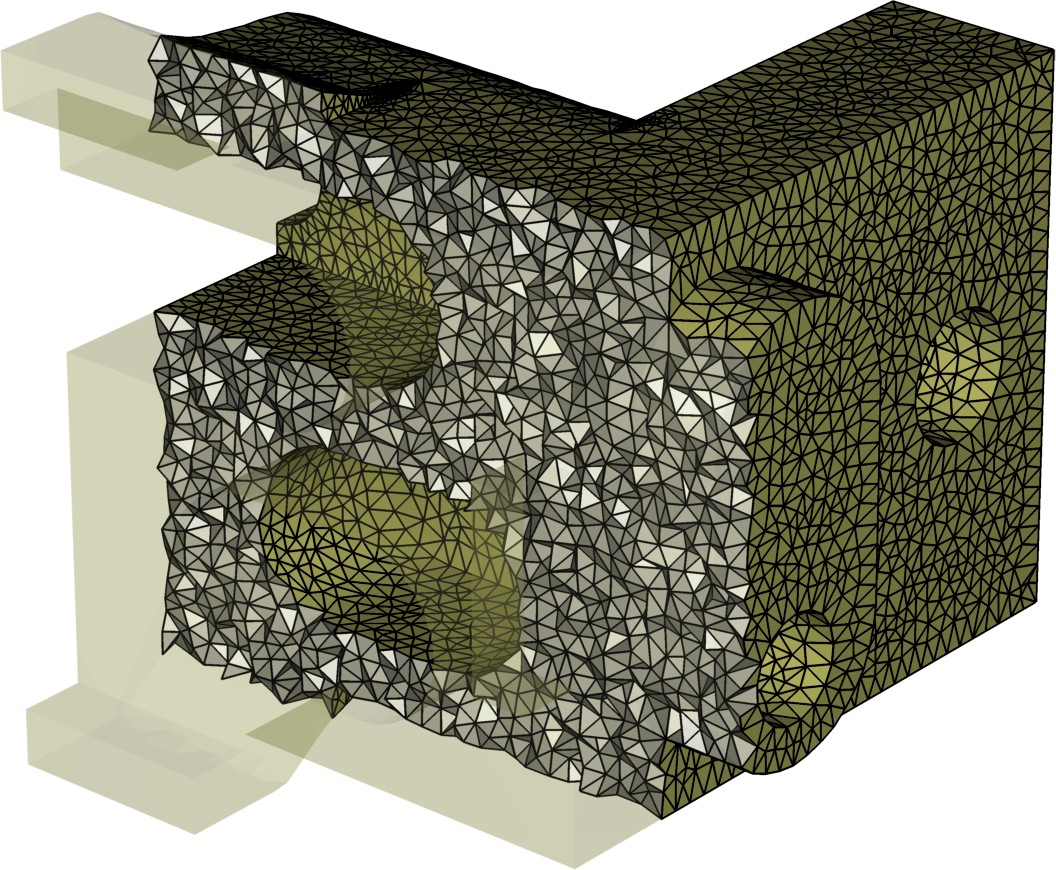} 
\end{center}
\end{minipage}

\\

\begin{minipage}[c]{0.31\textwidth}
\begin{center}
\includegraphics[width=4.90cm]{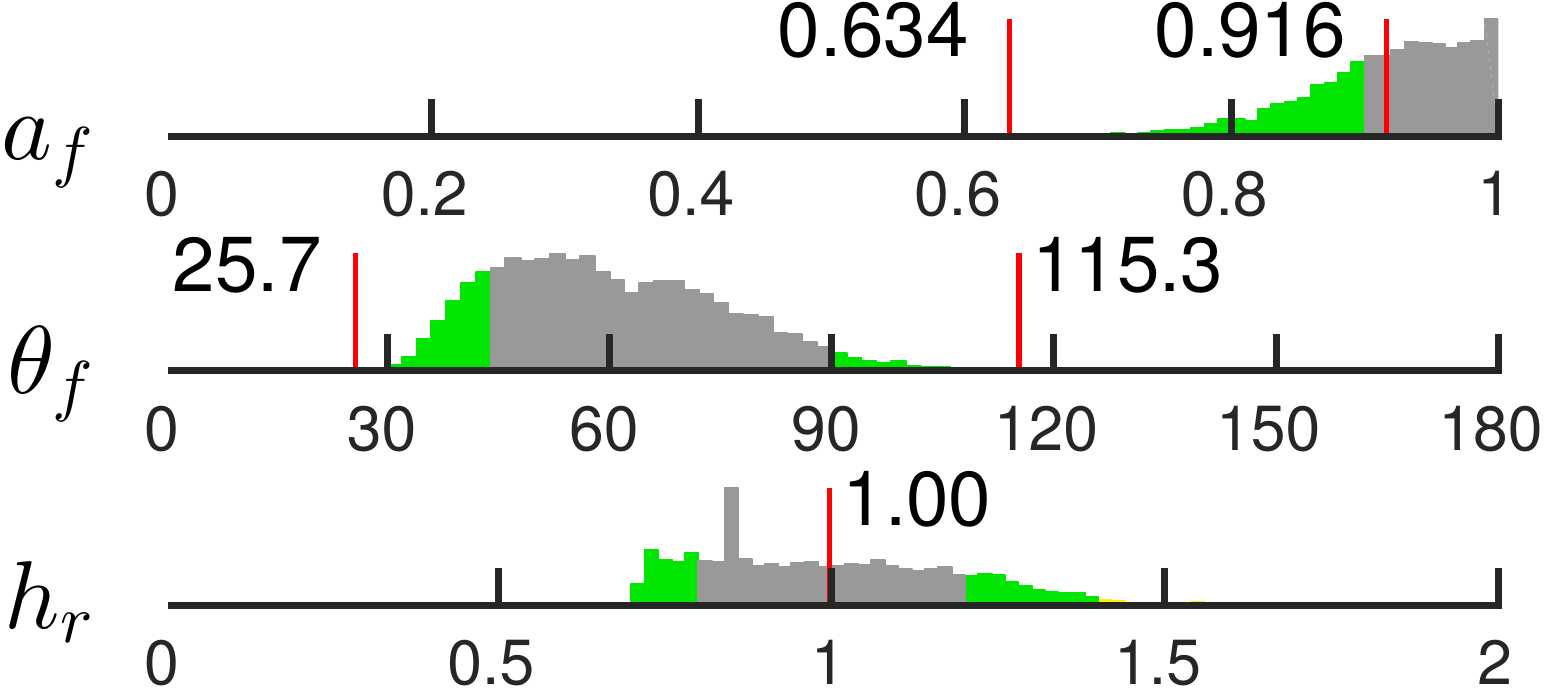} 
\end{center}
\end{minipage} &
\begin{minipage}[c]{0.31\textwidth}
\begin{center}
\includegraphics[width=4.90cm]{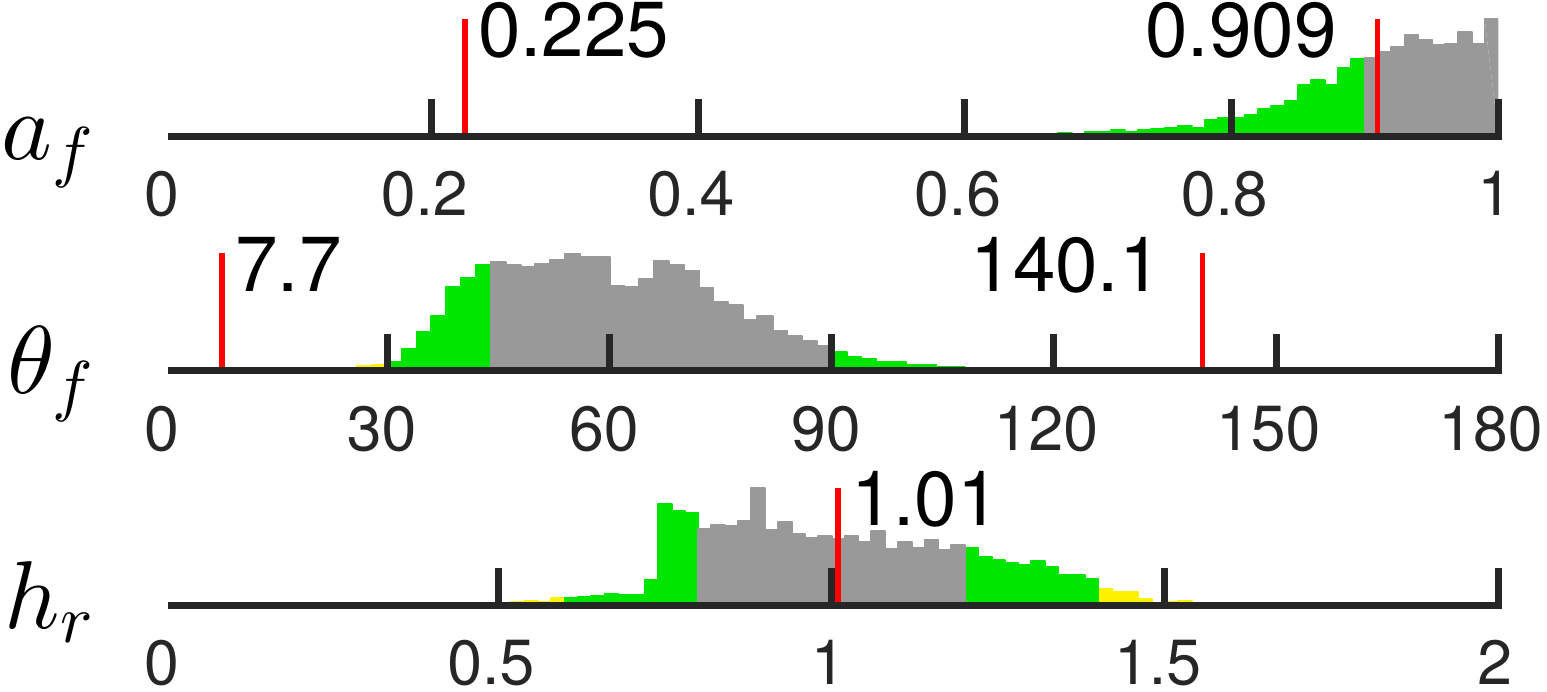} 
\end{center}
\end{minipage} &
\begin{minipage}[c]{0.31\textwidth}
\begin{center}
\includegraphics[width=4.90cm]{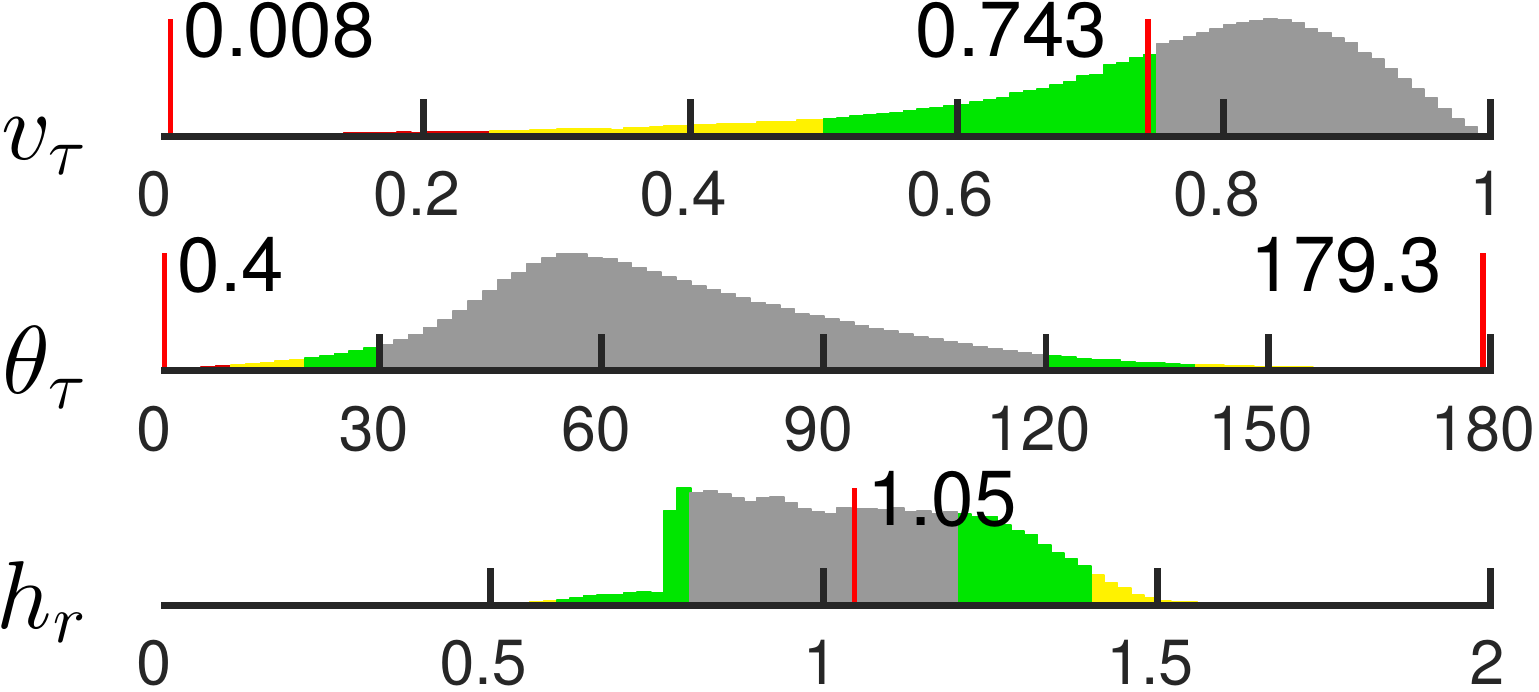} 
\end{center}
\end{minipage}

\end{tabu}}

\caption{Meshes for the \texttt{ISO}, \texttt{FANDISK} and \texttt{BRACKET} test-cases, showing output for the \texttt{JGSW-FD} and \texttt{CGAL-DR} algorithms. Detailed mesh statistics are shown including normalised histograms of element area-length and volume-length ratios, dihedral-angles and relative edge-lengths. Element counts and total refinement times are also shown.
}

\label{figure_meshes}

\end{figure}

\subsection{Preliminaries}

\medskip

The \texttt{JGSW-FD} and \texttt{CGAL-DR} algorithms were used to mesh a set of surface- and volume-based benchmark problems (Figure~\ref{figure_meshes}), comprising the \texttt{ISO}, \texttt{FANDISK} and \texttt{BRACKET} test-cases. The \texttt{ISO} test-case is an iso-surface problem, consisting of a multiply-connected collection of smooth surface patches with 1-dimensional constraints at the open boundary contours. The \texttt{FANDISK} object is a piecewise smooth surface model of a turbine component, incorporating a network of curve constraints inscribed at connections between adjacent surface patches. The \texttt{FANDISK} problem also includes sharply acute angle constraints, with a minimum angle of $20.1^{\circ}$ subtended by segments in the input geometry. The \texttt{BRACKET} object is a discrete CAD representation of a mechanical component, incorporating curve constraints at connections between adjacent surface patches, as per the \texttt{FANDISK} example. The \texttt{BRACKET} problem also incorporates acute angle constraints, with curve segments in the input geometry subtending angles as small as $6.5^{\circ}$. 

In all test cases, constant radius-edge ratio thresholds were specified for both surface and volume elements, such that $\bar{\rho}_{f} = 1.25$ and $\bar{\rho}_{\tau} = 2$, corresponding to $\theta_{\text{min}}\geq 23.5^\circ$ for surface facets. Additionally, uniform mesh-size and surface discretisation constraints were enforced, setting $\bar{h}(\mathbf{x}) = \alpha$ and $\bar{\epsilon}(\mathbf{x})=\beta \bar{h}\left(\mathbf{x}\right)$, with $\beta=\nicefrac{1}{4}$ and $\alpha$ a scalar length equivalent to approximately 3\% of the mean bounding-box dimension associated with each input geometry. The \texttt{JGSW-FD} and \texttt{CGAL-DR} algorithms impose mesh-size constraints in a slightly different manner, with \texttt{JGSW-FD} treating $\bar{h}(\mathbf{x})$ as a constraint on edge-length, and \texttt{CGAL-DR} treating $\bar{h}(\mathbf{x})$ as a constraint on the radii of the circumscribing balls associated with each edge, triangle or tetrahedron. To compensate for this difference, the mesh-size targets associated with triangles and tetrahedrons in \texttt{CGAL-DR} were reduced by a factor of $\nicefrac{4}{3}$. Note that such scaling ensures that \texttt{JGSW-FD} and \texttt{CGAL-DR} produce output with equivalent mean edge length metrics.  

For all test problems, detailed statistics on element quality are presented, including  histograms of element \textit{volume-length} and \textit{area-length} ratios $v(\tau)$ and $a(f)$, \textit{element-angles} $\theta(\tau)$ and $\theta(f)$, and \textit{relative-edge-length} $\bar{h}_{r}$. The element volume-length and area-length ratios are robust measures of element quality, where high-quality elements attain scores that approach unity. The relative edge-length is defined to be the ratio of the measured edge-length $\|\mathbf{e}\|$ to the target value $\bar{h}(\mathbf{x}_{e})$, where $\mathbf{x}_{e}$ is the edge midpoint. Relative edge-lengths close to unity indicate conformance to the mesh-size function. High-quality surface triangles  and interior tetrahedrons contain angles of $60.0^\circ$ and  $70.5^\circ$ respectively. %Histograms further highlight the minimum, maximum and mean values of the relevant distributions as appropriate.

\subsection{A Comparison of \texttt{JGSW-FD} and \texttt{CGAL-DR}}
\label{section_fdvscgal}

\medskip

The results in Figure~\ref{figure_meshes} show that, overall, the new \texttt{JGSW-FD} algorithm typically outperforms the \texttt{CGAL-DR} implementation -- generating slightly smaller meshes with improved element quality characteristics and mesh-size conformance. Overall computational expense for both algorithms was observed to be similar. In terms of element counts, the new method leads to a reduction of approximately 6\%. Focusing on the distributions of element shape-quality, it can be seen that the \texttt{JGSW-FD} algorithm achieves significant improvements in mean area-length and plane-angle distributions in the case of the \texttt{ISO} and \texttt{FANDISK} problems, with smaller improvements in the volume-length and dihedral-angle metrics realised in the \texttt{BRACKET} test-case. In all cases, it can be seen that tight, high-quality distributions ($a_{f}\simeq 1$, $v_{\tau}\simeq 1$ and $\theta_{f}\simeq 60^{\circ}$, $\theta_{\tau}\simeq 70^{\circ}$) are generated by the Frontal-Delaunay algorithm (\texttt{JGSW-FD}), while the standard Delaunay-refinement  approach (\texttt{CGAL-DR}) leads to broad, lower-quality distributions about similar means. Comparisons of distributions of element relative-length reveal the largest relative differences between algorithms, with the \texttt{JGSW-FD} implementation showing significantly improved conformance to the imposed mesh-size function, seen as a tight clustering of $\reledge\simeq 1$ in all test-cases. This result is not unexpected -- confirming that the new size-optimal off-centre point-placement scheme leads to high-quality vertex distributions that follow the imposed sizing function. These results are consistent with those previously obtained by the author for smooth manifold geometries \cite{Engwirda2016157,Engwirda2015330} using a simplified version of the three-dimensional Frontal-Delaunay-refinement algorithm presented here. These results demonstrate that the restricted Frontal-Delaunay paradigm can be extended to support piecewise smooth geometric inputs, including collections of curves, surfaces and enclosed volumes. 

Differences were also observed in the manner in which the \texttt{JGSW-FD} and \texttt{CGAL-DR} algorithms \textit{protect} sharp 1-dimensional features. Specifically, in the case of the \texttt{FANDISK} and \texttt{BRACKET} test problems, it was seen that the static \textit{protecting-balls} strategy utilised in the \texttt{CGAL-DR} algorithm led to significant local over-refinement, even creating elements with smaller angles than those imposed by the geometry itself. In contrast, the \textit{protecting-collar} approach developed in the current work, and deployed in the \texttt{JGSW-FD}, algorithm was observed to introduce a single isosceles element at the apex of any sharp 1-dimensional features, consistent with the methodology outlined in Section~\ref{section_protecting_angles}. In \texttt{JGSW-FD}, an associated local refinement of elements adjacent to these constraints is employed only to ensure that restricted edges are locally 1-manifold, typically leading to coarser local meshes. Overall, it is argued that the protecting-collar techniques developed in the current work offer a sparser and higher-quality solution to the problem of embedding arbitrary 1-dimensional constraints in restricted Delaunay meshes. Additional work focused on the development of high-quality protecting-collars for problems involving sharp surface features is currently in progress.

%\section{Conclusions \& future work}
%\label{section_conclusions}
%techniques to `protect' sharply acute surface features.

\bigskip

\textbf{Acknowledgements.} This work was conducted at the Massachusetts Institute of Technology and the University of Sydney with the support of a NASA--MIT cooperative agreement and an Australian Postgraduate Award. The author wishes to thank the anonymous reviewers for their helpful comments and feedback.

%\bibliography{references}
%\bibliographystyle{model1-num-names}

%\section*{References}

\bibliographystyle{elsarticle-num}
\bibliography{references}

\end{document}